# Size-Based Spectrophotometric Analysis of the Polana-Eulalia Complex

[1]McClure, L. T., [1]Emery, J. P.,[1]Thomas, C. A., [2]Walsh, K. J., [1]Williams, R. K.

*[1]Northern Arizona University, Department of Astronomy and Planetary Science, Flagstaff, AZ 86011, United States*

*[2]Southwest Research Institute, 1050 Walnut St. Suite 300, Boulder, CO, 80302, USA*

Correspondence Emails:
Lucas T. McClure – Email: ltm87@nau.edu
Joshua P. Emery – Email: Joshua.Emery@nau.edu
Cristina A. Thomas – Email: Cristina.Thomas@nau.edu
Kevin J. Walsh – Email: kwalsh@boulder.swri.edu
Riley K. Williams – Email: rkw48@nau.edu

**Abstract**: The Polana-Eulalia Complex (PEC) is an Inner Main Belt, C-complex asteroid population that may be the source of the near-Earth asteroid spacecraft mission targets (101955) Bennu and (162173) Ryugu. Here, we report a size-based investigation of the visible (VIS; 0.47 —0.89 μm) spectrophotometric slopes of the PEC's constituent families, the “New Polana” and Eulalia Families. Using two releases of the Sloan Digital Sky Survey's Moving Object Catalog as well as the 3rd data release of the Gaia catalog, we present evidence of size-based slope variability within each family. We find that Eulalia family members exhibit lower average slopes than Polana family members in all catalogs' samples, particularly for objects <9 km in diameter. We are unable to conclude that VIS slope distinguishability between the families is statistically significant, but we explore a potential cause of the bulk slope differences between the PEC families, in addition to providing commentary on size-slope trends generally.

## 1 Introduction

Asteroids relay numerous insights about the formation and evolution of the Solar System, as remnants from its formation and evolution. They are generally grouped spectrally (DeMeo et al., 2009) and dynamically (Nesvorný et al., 2015). Of the major spectral taxonomic classes, silicate-rich (S-complex) and carbon-rich (C-complex) asteroids dominate the mass in the Main Belt (MB, DeMeo and Carry, 2013). C-complex asteroids compose the majority of known asteroids, and they are also the predominant class of primitive asteroids, which are asteroids that have remained the least changed since their formation (*see*: DeMeo et al., 2009; Weisberg and Righter, 2014). Models of planetary migration during the early solar system suggest C-complex asteroids were delivered from the outer Solar System to the MB through gravitational interactions with migrating gas giants, Jupiter and Saturn (*e.g.,* Tsiganis et al., 2005; Gomes et al., 2005; Morbidelli et al., 2005; Walsh et al., 2011). Raymond and Izidoro (2017) has alternatively suggested that the delivery of C-complex materials occurred due to gas giant growth.

Gradie and Tedesco (1982) discovered the MB to have a clear silicate-to-carbon gradient with increasing heliocentric distance. DeMeo and Carry (2013) further demonstrated a similar stratification within the Main Belt, though they showed there to be significant mixing especially for the smallest asteroids. Despite the mixing, S-complex asteroids do comprise most of the mass of the Inner Main Belt (IMB; 2.1–2.5 AU), while primitive asteroids (mostly C-complex) dominate the mass of the Outer Main Belt (OMB; 2.82–3.3 AU) (DeMeo and Carry, 2013).

Asteroid families inform us about the dynamical evolution of the Solar System because they are dynamical clusters of asteroids (generally of the same taxonomic complex/type) residual from past collisions (Nesvorný et al., 2015). Characterizing physical characteristics present within families is helpful in tracing asteroid family members back to their parent bodies (Nesvorný et al., 2015; Masiero et al., 2015). The MB has numerous families near dynamical “escape hatches” to near-Earth space (*e.g.,* Knežević and Milani, 1991). Therefore, near-Earth asteroids (NEAs) can be linked dynamically back to MB (*e.g.*, Gladman et al., 2000), so spacecraft missions to NEAs have been a major focus over the last couple decades to compare NEA and MB populations.

The Polana-Eulalia Complex (PEC) is an IMB, C-complex asteroid population of substantial interest (Fig. 1). Comprised of the "New Polana" (hereafter, just Polana) and Eulalia families, the PEC exists near the $\nu_6$ secular resonance with Saturn that permits some of its members to enter near-Earth space (Knežević and Milani, 1991; Walsh et al., 2013). The targets of the *OSIRIS-REx* and *Hayabusa2* sample-return missions – (101955) Bennu and (162173) Ryugu, respectively – are two NEAs that have been connected to the PEC due to orbital evolution models (Campins et al., 2010, Campins et al., 2013; Bottke et al., 2015) and spectral matching (de León et al., 2016; Pinilla-Alonso et al., 2016). There is, however, spectral discrepancy between the PEC and Bennu (Pinilla-Alonso et al., 2016). Bennu exhibits a globally negative (blue) near-infrared (NIR; 0.9–2.2 µm)spectral slope (Clark et al., 2011; Hamilton et al., 2019), whereas the PEC and Ryugu display positive (red) NIR spectral slopes (Pinilla-Alonso et al., 2016; Kitazato et al., 2019). However, the returned Bennu sample shows a red spectral slope in the visible (VIS; 0.47 – 0.89 µm) to NIR range, suggesting spectral differences with depth (Lauretta et al., 2024). Nonetheless, the cause of Bennu's blue (global) NIR spectral slope in the context of the red NIR spectral slopes of its likely origin remains puzzling.

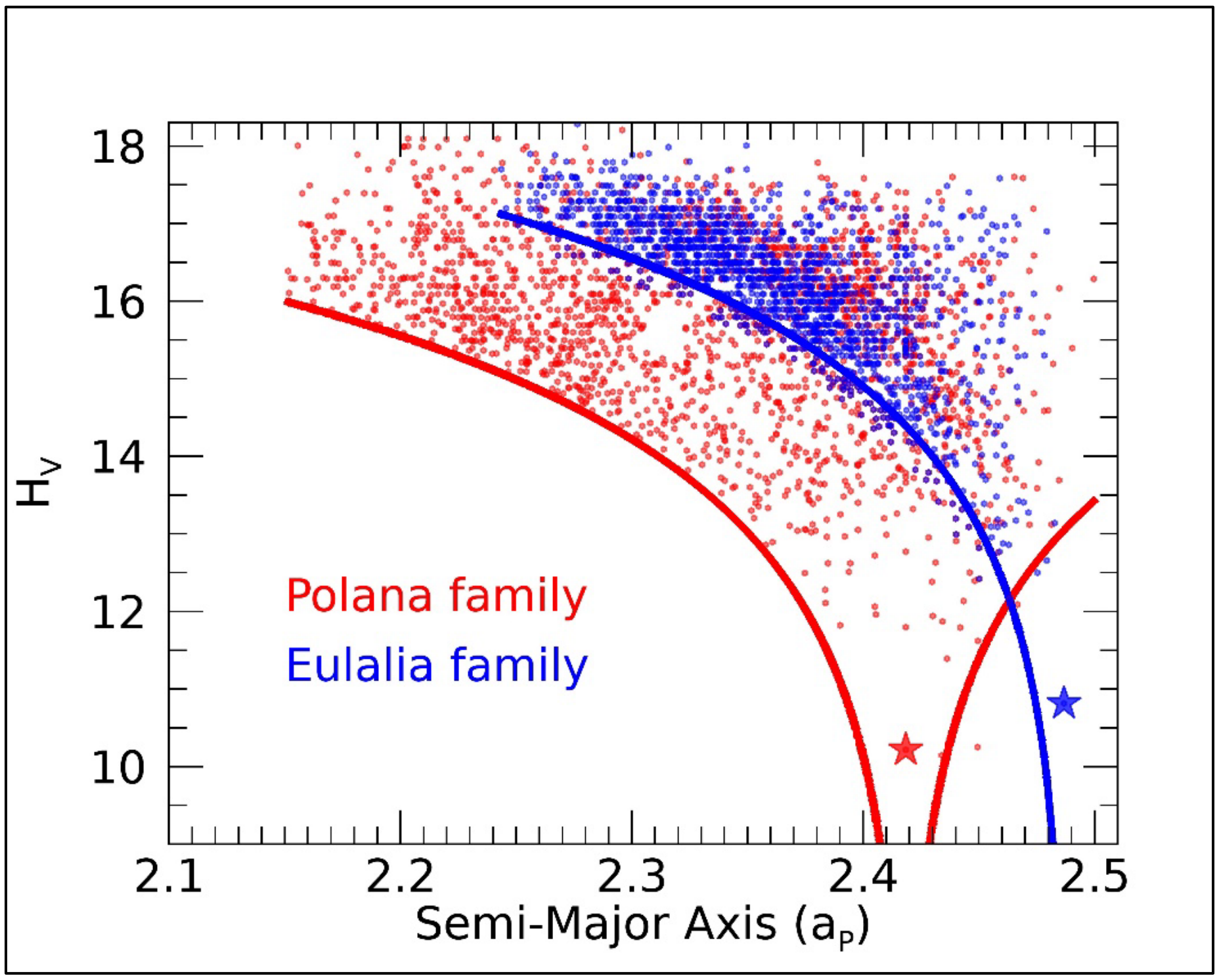

**Figure 1**: *The "V"-shaped spreads of the Polana (red) and Eulalia (blue) family members that are caused by the "Yarkovsky drifts" of their members' orbits (color-corresponding curves). Largest members, (142) Polana and (495) Eulalia, are represented by the red and blue stars, respectively.*

The PEC families can only partially be distinguished through their dynamical properties. The Polana family is centered about (142) Polana at a semi-major axis, a, of 2.418 AU, and the Eulalia family is centered about (495) Eulalia at roughly a = 2.488 AU (see Fig. 1 ). Being the older family of the PEC, the Polana family's “V”-shaped spread (from Yarkovfsky Drift; Vokrouhlický et al., 2006 ; Walsh et al., 2013 ) is considerably wider than that of the Eulalia family, but the two families have substantial overlap in a, eccentricity (e), and inclination (i) (Walsh et al., 2013).

The PEC families may be spectrally separable, but the literature has presented varied results. Ground-based, spectral surveys of the PEC were not able to distinguish the families spectrally in either VIS (de León et al., 2016 ) or NIR ( Pinilla-Alonso et al., 2016 ) wavelengths. Tatsumi et al. (2022) used Eight Color Asteroid Survey (ECAS) data to investigate the PEC at near-ultraviolet to visible (NUV-VIS; 0.35–0.95 μm) wavelengths, further reporting indistinguishability. However, Delbo et al. (2023) used the 3rd data release of the Gaia catalog (GDR3; Galluccio et al., 2023 ) to show the Polana and Eulalia families are spectrally distinguishable at NUV wavelengths (0.35–0.55 $\mu m^{-1}$). Delbo et al. (2023) also reanalyzed the PEC samples from Tatsumi et al. (2022) and found that the families are actually significantly distinguishable, underscoring the necessity to probe these two families more finely at VIS (and eventually longer) wavelengths with larger datasets.

Several properties have been observed to vary with asteroid size, with asteroidal characteristics generally changing around 10 km (e.g., Binzel et al., 2004 ; Vokrouhlický et al., 2006 ; Thomas et al., 2012 ; Vokrouhlický et al., 2015 ; Walsh, 2018 ; DellaGiustina et al., 2019 ; Sugita et al., 2019 ; Morate et al., 2019 ; Thomas et al., 2021 ; Beck and Poch, 2021 ; MacLennan and Emery, 2022 ). C-complex populations show size-dependent, VIS spectral slope trends (hereafter, “size-slope” trends; see: Morate et al., 2019 ; Beck and Poch, 2021 ; Thomas et al., 2021 ). Morate et al. (2019) demonstrated that IMB families of C-complex asteroids exhibit increasing variability in VIS spectral slopes as absolute magnitude (HV) increases (i.e., size decreases), generating a “cone” shape (hereafter, “Morate Cone”). However, Tinaut-Ruano et al. (2024) showed that the Morate Cone structure is not a bulk property of the MB. Thomas et al. (2021) used spectrophotometric data from 4th Release of the Sloan Digital Sky Survey's Moving Object Catalog (SDSS MOC4) to show that 9 MB families of C-complex objects exhibit size-slope trends that bifurcate into “types.” The “Hygiea-type” is characterized by bluing then reddening from small to large sizes, whereas the “Themis-type” shows the opposite behavior.

Previous works (e.g., Morate et al., 2019 ; Thomas et al., 2021 ) have suggested a variety of potential factors to explain the size-slope behaviors of C-complex populations. Morate et al. (2019) suggested that collisional history and relatively rapid surface evolution of smaller members may allow for greater spectral slope variability compared to their larger counterparts. In suggesting causes for the size-slope trend types observed by Thomas et al. (2021) , they discussed compositional and grain size trends may play roles in the spectral diversity within families. However, the cause or causes of size-slope trends currently remain unknown.

Size-based spectral analyses of the PEC may provide a quantitative means to separate the PEC families in VIS wavelengths and for various sizes. Ground-based ( de León et al., 2016 ) and catalog-based ( Delbo et al., 2023 ) works did not find significant VIS differences between the PEC families nor did they find any size-slope relationships for the PEC. However, de León et al. (2016) was restricted by sample size and Delbo et al. (2023) did not focus solely on the PEC families. As a C-complex population, the PEC may exhibit size-slope behavior consistent with prior literature (i.e., Morate et al., 2019 ; Thomas et al., 2021 ), and size-slope analysis for the PEC could provide a way to compare the slopes of these two families at various sizes. Thus, we hypothesize differences in size-slope behaviors of the Polana and Eulalia families. We further expect the Eulalia family will be spectrally and significantly bluer than the Polana family at VIS wavelengths, in agreement with the result from Delbo et al. (2023).

# 2 Methods

## 2.1 Data Sources & Acquisition

We searched the ~100,000 moving objects in the SDSS MOC4 to identify cataloged members of the PEC families (Walsh et al., 2013). For each selected object, we extracted magnitudes in the g' (λg' = 0.4686 µm), r' (λr' = 0.6166 µm), i' (λi' = 0.7480 µm), and z' (λz' = 0.8932 µm) filters. We used MOC4 to maintain consistency with and make direct comparisons to Thomas et al. (2021) (see: Section 4.5).

We also broadened our data sources for spectrophotometry of PEC members to both include newer datasets as well as mitigate any dataset dependencies in our results. We extracted g', r', i', and z' magnitudes from the SDSS MOC release made by Sergeyev and Carry (2021) (hereafter, S&C21). S&C21 scanned images from the SSDS that occurred after the 4th Release, used looser identification criteria (*e.g.,* no apparent velocity limit), and included larger sky coverage, resulting in >1 million observations that include ~380,000 known moving objects in their catalog. S&C21 also includes probabilistic taxonomic information for a portion of the cataloged objects. We also used the GDR3, which overlaps the wavelength range of the SDSS MOC releases (Galluccio et al., 2023). We collected reflectance measurements from filters that span roughly the same range as the SDSS MOC releases (*i.e.,* the filters from 0.462 to 0.902 µm).

We identified PEC members in each catalog based on asteroid number and/or designation. The family lists for the PEC record asteroid numbers, which were used for identification within the SDSS MOC releases. Sometimes, the SDSS MOC releases list objects only by designations. In these instances, we used the Python HorizonClass interface within astroquery to search the JPL Horizons database for all designations associated with PEC member numbers provided in the family lists. Objects in the SDSS MOC releases that have neither a recorded designation nor a number were not included in the acquisition process, since data parsing on the level of matching orbital dynamics to currently-known objects was outside this work's scope. All asteroids reported in GDR3 are listed by asteroid number.

We enforced error and limiting-magnitude cutoffs for each identified object in each of the SDSS MOC releases. For both SDSS MOC releases, we selected objects with errors in the *g'*, *r'*, *i'*, and *z'* filters that are <6 % of the magnitude to balance data acquisition with reliability (Thomas et al., 2021). Objects were also selected only if their magnitudes did not exceed the limiting magnitudes for the *g'*, *r'*, *i'* and *z'* filters (DeMeo and Carry, 2013; Thomas et al., 2021). Objects with flags that indicate additional, non-photometric conditions were also not selected (DeMeo and Carry, 2013; Thomas et al., 2021). Different flags are provided for the photometric measurements in S&C21, so we selected objects without any flags associated with the *g'*, *r'*, *i'*, and *z'* filters in that SDSS MOC release. For GDR3 samples, we only selected objects without a flag in any of the filters used. We slightly raised the error cutoff to <8 % of the reflectance for each GDR3 band used so as to retrieve an adequate amount of small (< 9-km) objects.

We also selected objects not flagged as being potential dynamical interlopers, based on the ratio described in C parameter space (*see:* Nesvorný et al., 2015).

$$C = |C_J / C_0|, \quad \textbf{(Eq. 1)}$$

where $C_J$ is calculated from each (potential) family member's semi-major axis difference from the family center as well as $H_V$ (Nesvorný *et al.,* 2015). The central semi-major axes, $a_c$, of the constituent families that we used are $a_c$ = 2.418 AU and $a_c$ = 2.448 AU for the Polana and Eulalia families, respectively (i.e., semi-major axis of the largest member of each family from Walsh *et al.* 2013). $C_0$ is a constant (for each family) calculated from the width of the size-dependent, "V"-shaped spread of a family in semi-major axis space (Nesvorný *et al.,* 2015). We used $C_0$ values reported in Walsh et al. (2013) for the Polana and Eulalia families ($C_{0,Polana}$ = 0.000169 and $C_{0,Eulalia}$ = 0.000092, respectively). Objects with $|C| > 1$ are defined as potential dynamical interlopers that are outside of the "V"-shape spread for their family for their particular $H_V$. Only the Eulalia family was affected by this dynamical criterion.

All extracted objects were further selected on a basis of visual geometric albedo, $p_v$, as reported by WISE and cataloged in JPL's Small-Body Database Lookup (Mainzer et al., 2016). We selected objects that exhibit reported albedos of $p_v$ < 12%, a threshold for C-complex objects that was identified in Delbo *et al.* (2017) and used in Delbo *et al.* (2023). The Polana family was mostly affected by this criterion, likely because of S-type interlopers from the Nysa family (*see:* Delbo *et al.* 2023).

We also selected objects based on colors as an additional way to avoid inclusion of Nysa interlopers. For SDSS MOC releases, we selected PEC members with $a^* < 0$, where a* is defined by (**Eq. 2**):

$$a^* = 0.89(g' - r') + 0.45(r' - i') - 0.57 \quad \textbf{(Eq. 2)}$$

The a* parameter is the first principal component of a SDSS-based color-color plot, roughly separating C-complex (a* < 0) from S-complex (a* > 0) asteroids (Ivezić *et al.,* 2001). Additionally, objects from S&C21 were selected if they exhibit a probability > 0.5 of being a class of primitive asteroid or if they met all other selection criteria yet were "unclassified." The SDSS MOC releases include objects with numerous observations, so if the average a* value for all observations was < 0, then we include the object in our sample, particularly since the observations met all aforementioned criteria. For GDR3 data, we used a color criterion from Delbo *et al.* (2023) that separates these taxonomic classes: $-0.2 < R_z - R_i < 0.185$, where $R_i$ is the reflectance at 0.7480 μm and $R_z$ is the reflectance at 0.8932 μm. Since Gaia does not have filters centered on those specific wavelengths, we generated a spline between the filters centered on 0.726 μm and 0.902 μm in order to calculate $R_z - R_i$ color. While the Polana family was more affected by this criterion (again, likely due to Nysa family interlopers), both families were affected.

## 2.2 Slope Determination, Selection, & Usage

For each object in each sample, we calculated spectrophotometric slope. To remove solar contributions from each SDSS-sourced object, we subtracted 0.44, 0.11, and 0.03 from (g'-r'), (r'-i'), and (i'-z') colors, respectively. Solar-corrected colors were then used to convert to reflectance values, normalized to the reflectance in the *r'* filter. For consistency, we re-normalized GDR3 data to the SDSS's r' filter wavelength ($\lambda_{r'}$ = 0.6166 μm) by fitting a line between the Gaia filters centered around λ = 0.594 μm and λ = 0.638 μm to derive a reflectance value at $\lambda_{r'}$. Slopes were computed by a least-squares fit to the reflectance data. The slopes from multiple SDSS observations of a single object were averaged. Calculated slopes are reported in units of reflectance per μm ($\mu m^{-1}$). Slope error was propagated from the errors in each of the filters used. We did not perform phase angle corrections to our slope values presented in this work, since phase reddening coefficients of C-complex asteroids are negligible compared to our calculated slopes and their uncertainties. Lantz et al. (2018) determined phase reddening coefficients for suites of C-, Ch-, and B-types and found each type's coefficient to be on the order of $10^{-3}$ $\mu m^{-1}$ $deg.^{-1}$. Phase reddening effects on specific objects, like Bennu ($4.4 \times 10^{-3}$ $\mu m^{-1}$ $deg.^{-1}$; Fornasier et al., 2020) and Ryugu ($2.0 \times 10^{-3}$ $\mu m^{-1}$ $deg.^{-1}$; Tatsumi et al., 2020), also align with this order of magnitude. We conducted a test using the Lantz et al. (2018) phase reddening coefficient on all objects in each sample, and this correction negligibly affected the observed trends in **Section 3**. Corrections to Hyades 64 for the blue photometer were applied for our GDR3-sourced objects, affecting only the filters centered on 0.462 μm and 0.506 μm (Tinaut-Ruano et al., 2023).

We imposed a slope cutoff to only include objects with slopes < $|1\ \mu m^{-1}|$ because a handful of "very blue" or "very red" Polana family members made it through the aforementioned criteria in **Section 2.1**. Delbo et al. (2023) only included GDR3-sourced slopes from $-0.5\ \mu m^{-1} < S' < 0.6\ \mu m^{-1}$, while Thomas et al. (2021) did not impose slope cutoffs for their SDSS-based work. We did not seek to limit data acquisition too strictly, but we wanted to acknowledge that the reliability of objects with high (SDSS-reported) slopes has been called into question by Humes (2023).

We applied three kinds of investigations to probe slope distributions at various sizes. Firstly, we surveyed bulk slope distributions for both families in each catalog to compare average slope and standard deviation (σ). Secondly, we divided each bulk sample into four, equal-sized bins in $H_V$-space to investigate variability within each size bin. Slope mean, slope σ, and standard deviation of the slope mean ($\sigma_m$) were calculated to quantify the spread of the PEC families in each size bin. Thirdly, we performed a running box mean of $H_V$ (and WISE-derived diameter) and slope to investigate size-slope behavior, similar to the method in Thomas et al. (2021). All objects in this study have WISE-derived diameters (D), gathered from JPL's Small-Body Database Lookup (see: Mainzer et al., 2016). Errors in the running box mean were calculated with the standard deviation of the binned slopes as well as the propagated errors within each bin. The bin size was set to be 20 % of the sample size – a percentage that allowed for the ability to observe small-scale changes in slope averages in the samples (*see*: Binzel et al., 2004; Thomas et al., 2012; Thomas et al., 2021). Binned slopes were then compared to sorted, binned size parameters (*i.e.,* $H_V$ and D).

### 2.3 Selected Samples

We obtained 6 samples with adequate sample size (*see:* Table 1) – the Polana and Eulalia families for each of the three catalogs. SDSS samples are roughly congruent with the size frequency distribution (SFD) of their respective family. The S&C21 samples have ~1.5 times as many objects as our MOC4 samples. GDR3-derived samples contain relatively fewer very small (< 5 km) asteroids and more very large (> 15-km) objects than the SDSS samples, so their SFD turns over at larger sizes than the SDSS samples.

Despite inclusion of large members, over ¾ of the GDR3 samples have diameters <9 km (Mainzer et al., 2016; *see:* **Section 2.1** and Table 1).

# 3 Results

## 3.1 Bulk Slope Distributions

The Polana family's average VIS slope is more positively-sloped (*i.e.,* redder) than the Eulalia family's (Table 1 and Fig. 2). We particularly find the Polana family is redder than the Eulalia family at small sizes ($H_V > 14$ or $D < 9$ km; Fig. 3, Fig. 4). Slope distributions of the two PEC families are statistically distinguishable in the SDSS MOC samples, according to Kolmogorov–Smirnov (K—S) tests (Table 1). Although the K—S statistic for the <9-km GDR3 subsample is ~2.6× smaller than the K—S statistic for the bulk GDR3 sample, both indicate that the two families are not significantly distinguishable in VIS wavelengths (Table 1).

The Polana family exhibits a wider spread of spectrophotometric slope than the Eulalia family (*i.e.*, dashed lines in Fig. 2). In SDSS samples, the Polana family has a slope spread of roughly $-0.7\ \mu m^{-1} < S' < 0.7\ \mu m^{-1}$ ($\sigma = 0.17$), whereas the Eulalia family exhibits a tighter spread from roughly $-0.5\ \mu m^{-1} < S' < 0.5\ \mu m^{-1}$ ($\sigma = 0.13-0.14$). The GDR3 samples display the same relative nature of the spreads, though over redder slope ranges. That is, GDR3-derived samples show much redder average slopes than the SDSS samples (Table 1), underscored by GDR3 slope distributions of the Polana and Eulalia families being $-0.5\ \mu m^{-1} < S' < 1\ \mu m^{-1}$ ($\sigma = 0.19$) and $-0.3\ \mu m^{-1} < S' < 0.5\ \mu m^{-1}$ ($\sigma = 0.16$), respectively. Additionally, many of the small ($H_V > 14$ or $D < 10.5$ km) Eulalia family members cluster within the bluer portion of the wider Polana family slope distribution.

Polana family samples show a “tail” of positively-sloped objects beyond +2σ, especially in the GDR3 sample. These objects reside in the smaller-sized portion of the sample ($H_V > 13.5$ or < 8 km). Notably, the tail of redder objects is also present in the GDR3 Eulalia family sample, though the tail is not as red as the Polana family’s (Fig. 2c).

**Table 1**: The means (with standard deviations of the means), standard deviations, and K-S probability for each sample in this work.

| Dataset | Family | Sample Size | Mean (+/- $\sigma_m$) [$\mu m^{-1}$] | σ [$\mu m^{-1}$] | K-S Statistic |
|---|---|---|---|---|---|
| SDSS MOC (4th Release) | Polana | 236 | -0.017 (+/- 0.0111) | +/- 0.17 | 0.019 |
| | Eulalia | 145 | -0.076 (+/- 0.0107) | +/- 0.13 | |
| SDSS MOC (S&C21) | Polana | 355 | -0.015 (+/- 0.00927) | +/- 0.17 | 0.055 |
| | Eulalia | 225 | -0.057 (+/- 0.00959) | +/- 0.14 | |
| Gaia Data Release 3 (all objects) | Polana | 361 | 0.096 (+/- 0.00986) | +/- 0.19 | 0.726 |
| | Eulalia | 166 | 0.077 (+/- 0.012) | +/- 0.16 | |
| Gaia Data Release 3 (< 9-km objects) | Polana | 273 (76%) * | 0.13 (+/- 0.0121) | +/- 0.20 | 0.270 |
| | Eulalia | 137 (83%) * | 0.094 (+/- 0.0141) | +/- 0.17 | |

**Percentages with respect to the bulk GDR3 samples.*

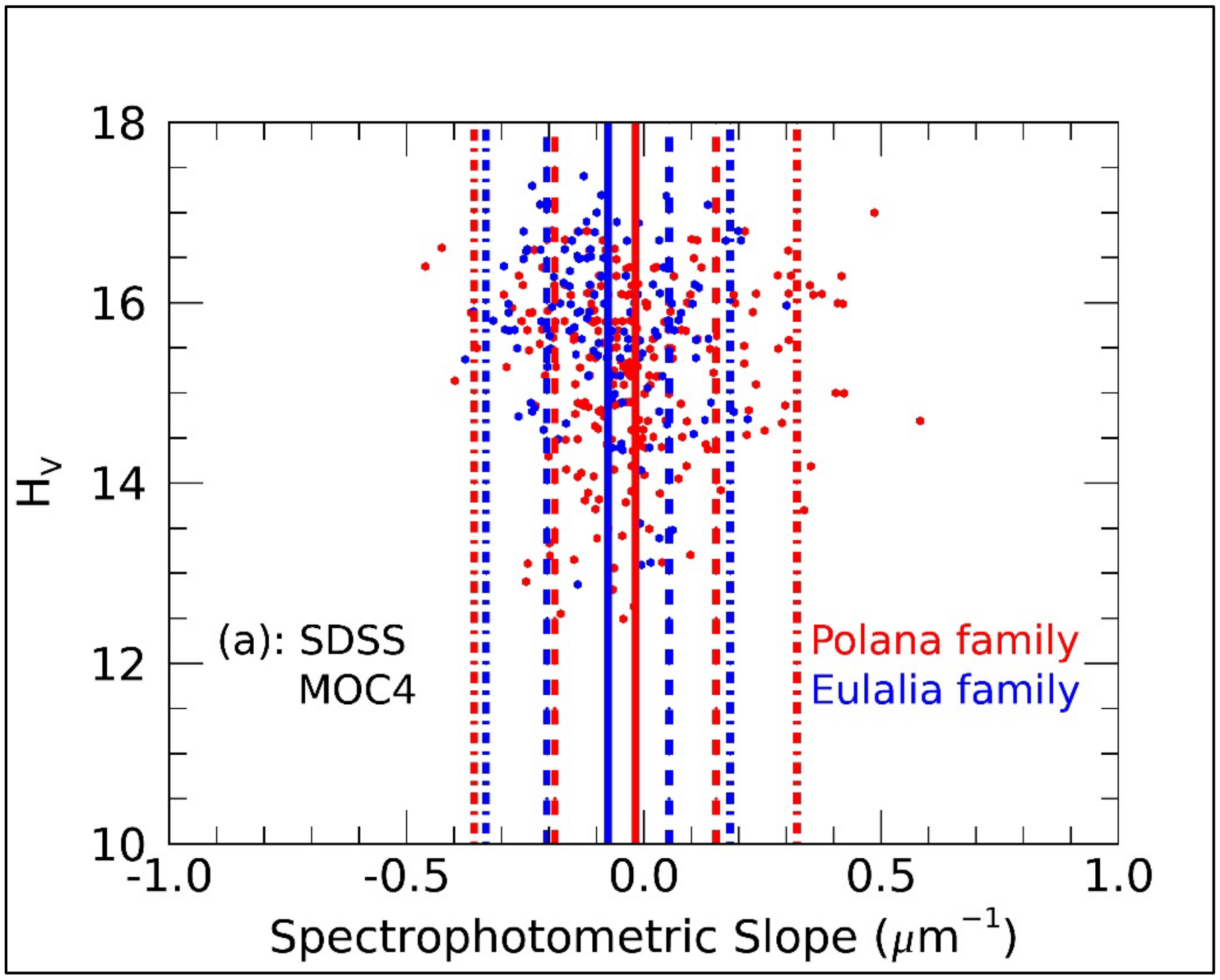

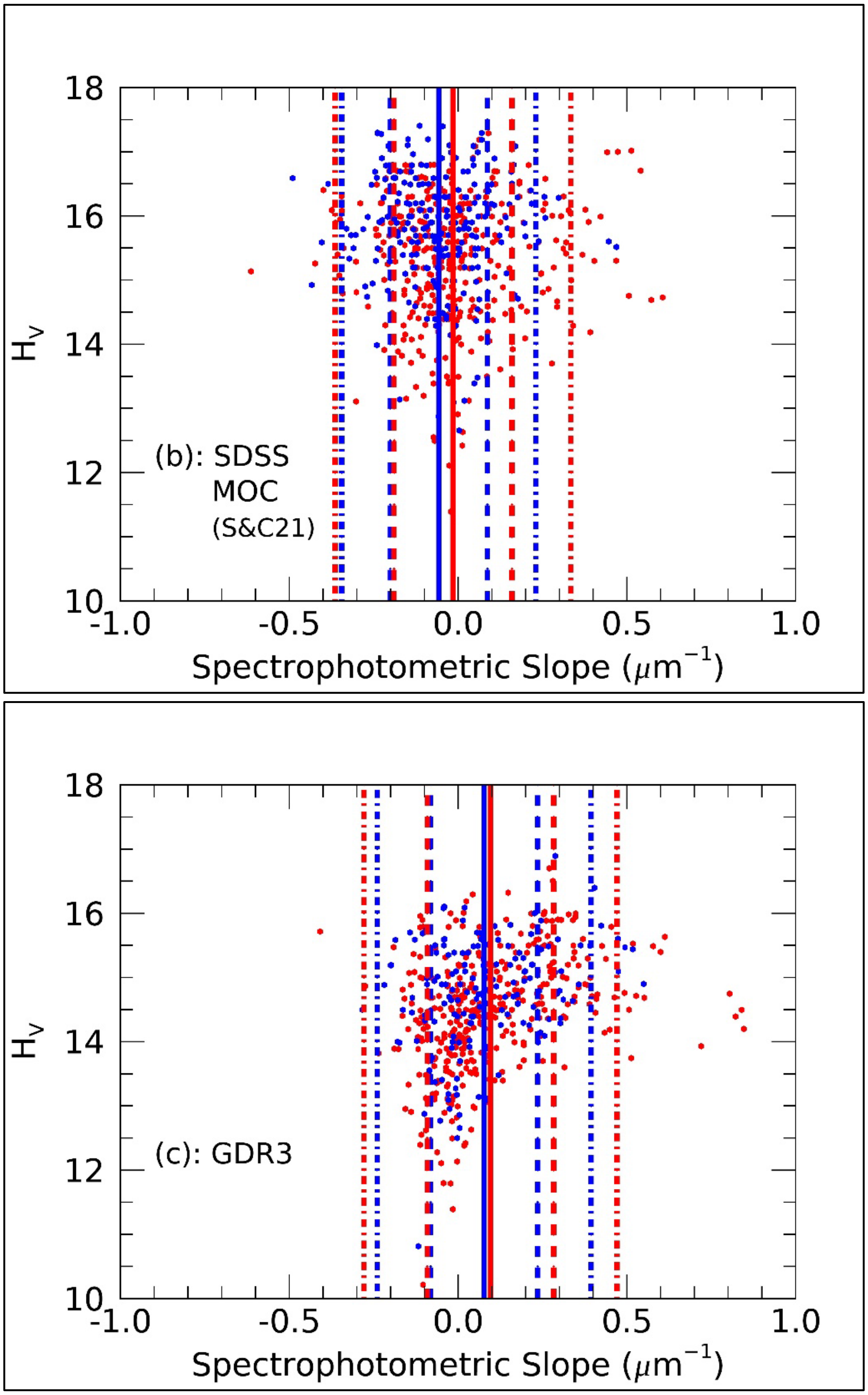

**Figure 2**: *The bulk distributions of visible spectrophotometric slope for the Polana (red) and Eulalia (blue) families in each of the three datasets used in this work:* ***(a)*** *SDSS MOC4,* ***(b)*** *SDSS MOC from S&C21, and* ***(c)*** *the 3rd Data Release of Gaia. The solid, color-correspondent vertical line represents the average slope for the respective PEC family. Color-coded dashed and dash-dotted lines represent ±1σ and ±2σ standard deviations, respectively.*

**3.2 Size-Slope Variability**

To investigate the size-dependence of slope distributions expressed in Fig. 1 and Table 1, we divide each of the bulk distributions into 4 bins of $H_V$ with increment determined by $(H_{V,Max} - H_{V,Min}) / 4$, where $H_{V,Max}$ and $H_{V,Min}$ are the maximum and minimum $H_V$ in each sample, respectively. As shown in Fig. 3 and Table 2, both families show some slope variability with size in each dataset. In the case of the MOC4 samples (Fig. 3a), both families have a fluctuating range of slopes in their four bins. For the Polana family's S&C21 sample, increasing variability with decreasing size is exhibited (Fig. 3b). The Eulalia family's S&C21 sample, however, shows consistent slope ranges at each size bin, with the exception of the bin of its largest members (Fig. 3b). In the GDR3 samples, both families show redder average slopes as size decreases. Slopes span a wider range for the Polana family's smaller members (*i.e.,* the 2 bins containing smallest members), considerably wider than slope ranges for larger members (*i.e.,* the 2 bins containing largest members; Fig. 3c). The GDR3 size-slope pattern for the Eulalia family is similar to its SDSS MOC counterparts. Partition of the bulk sample into exactly 4 bins often generates small subsample sizes in some bins, which cannot be used to adequately constrain size-slope variability (Table 2).

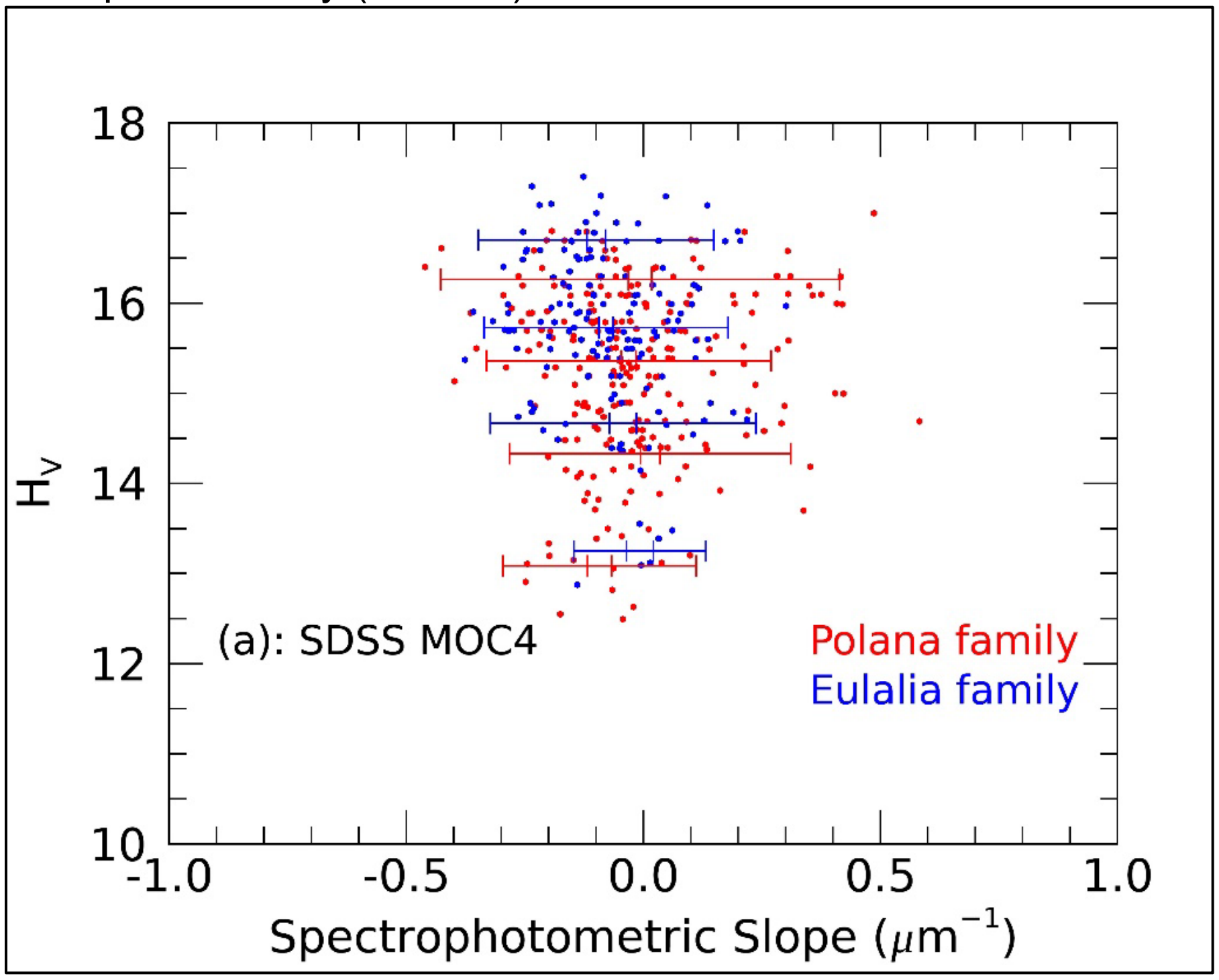

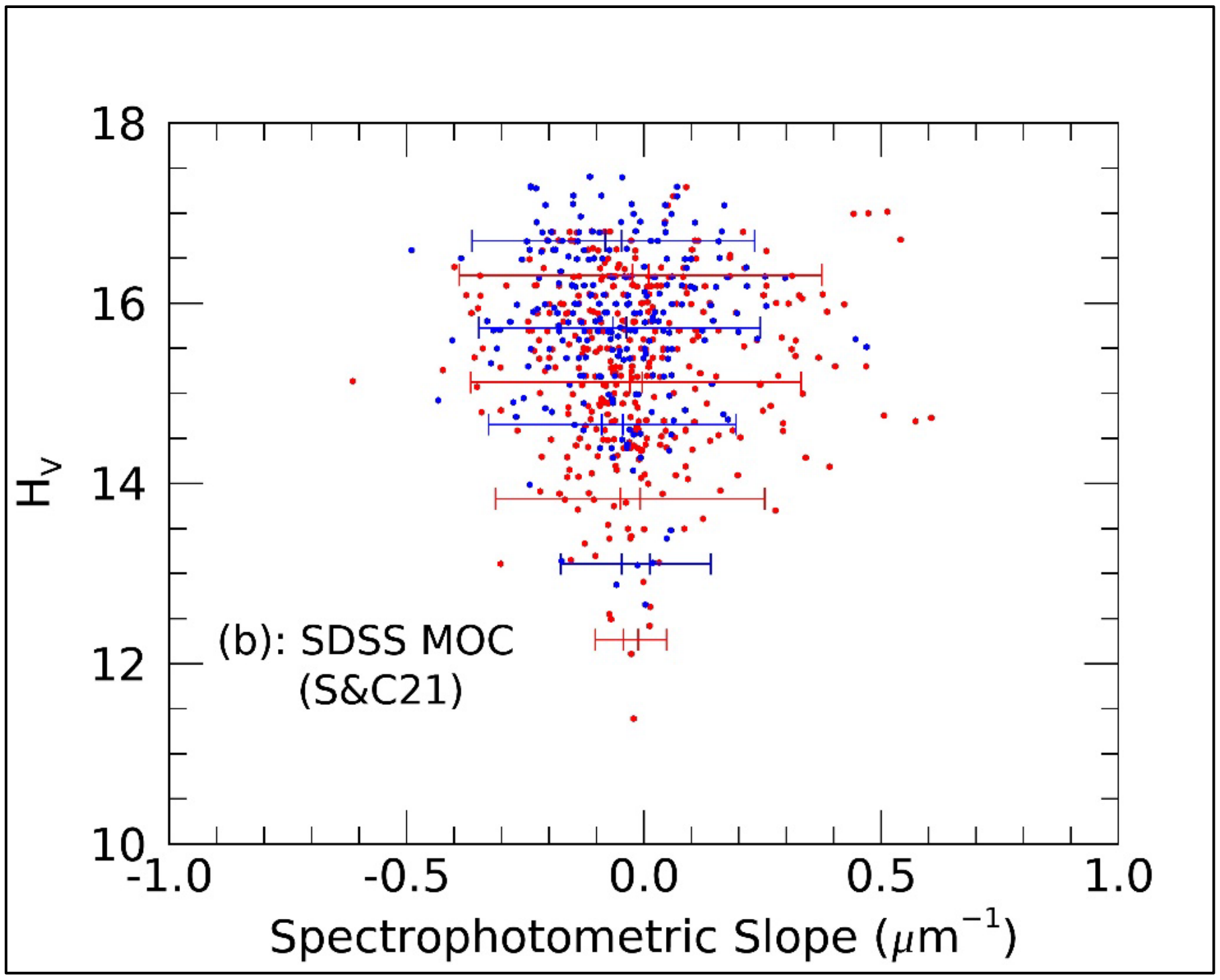

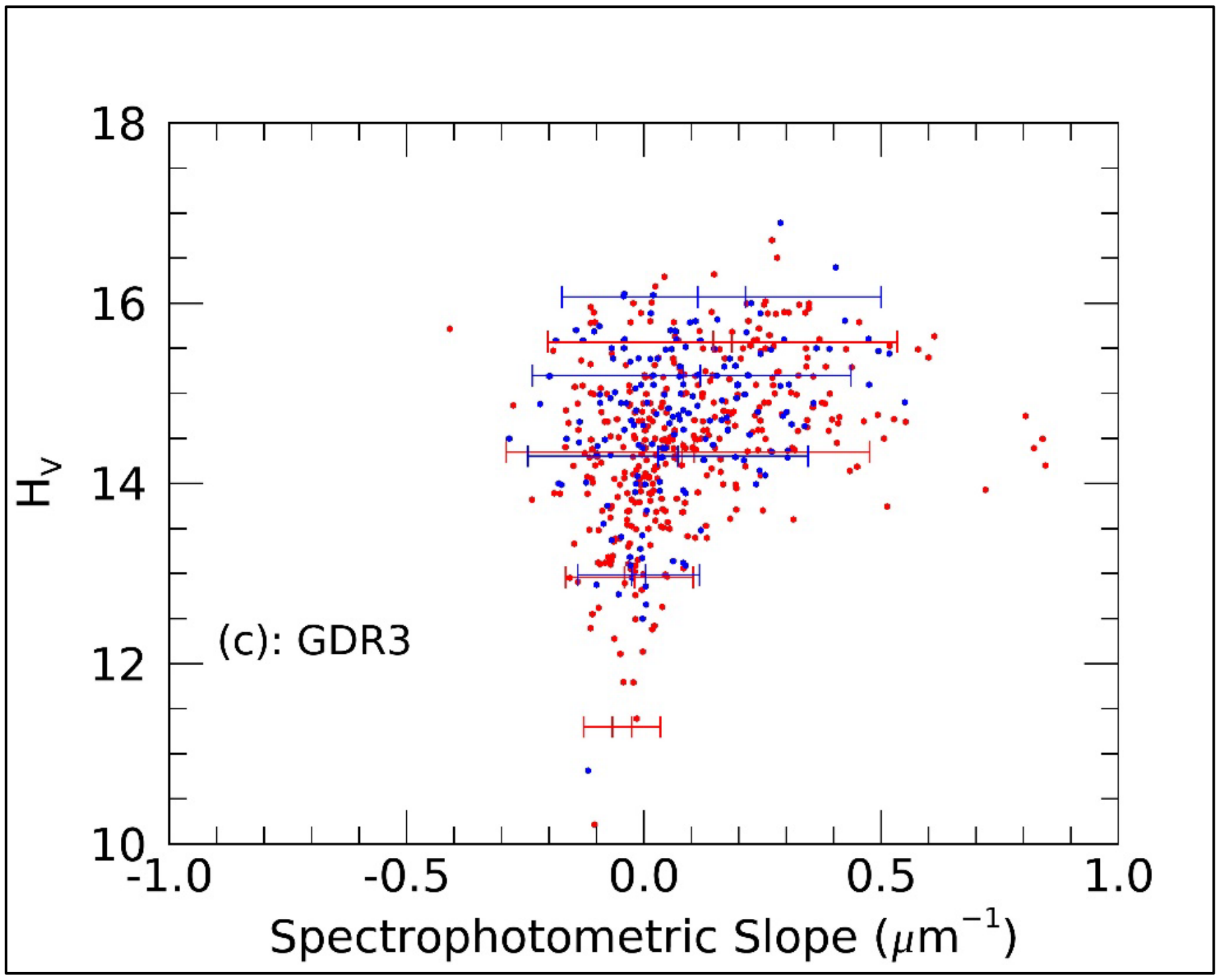

**Figure 3:** *Both the Polana (red) and Eulalia (blue) families are divided into 4 bins determined by their $H_{Min}$ and $H_{Max}$ values. The mean slope and size, the standard deviation of the slopes, and slopes' standard deviations of the mean are calculated for each subsample. The outer tick marks represent ±2σ and the inner tick marks represent*

*$\sigma_m$. The process is conducted and shown for the **(a)** SDSS MOC4, **(b)** SDSS MOC from S&C21, and **(c)** GDR3.*

**Table 2**: The mean, median, and standard deviation of the sample (σ) and of the mean ($\sigma_m$) are provided for each of the $H_V$ bins. The 4 equal-sized bin sizes are determined by the minimum and maximum $H_V$ in each sample.

| Dataset | Family | Average $H_V$ | N | Mean Slope ($\mu m^{-1}$) | Median Slope ($\mu m^{-1}$) | σ ($\mu m^{-1}$) | $\sigma_m$ ($\mu m^{-1}$) |
|---|---|---|---|---|---|---|---|
| SDSS MOC4 | Polana | 13.08 | 16 | -0.092 | -0.066 | 0.10 | 0.025 |
| | | 14.33 | 52 | 0.014 | -0.012 | 0.15 | 0.021 |
| | | 15.36 | 93 | **-0.031** | -0.051 | 0.15 | 0.016 |
| | | 16.26 | 75 | -0.007 | -0.047 | 0.21 | 0.024 |
| | Eulalia | 13.25 | 6 | -0.007 | 0.015 | 0.070 | 0.028 |
| | | 14.67 | 24 | -0.043 | -0.045 | 0.14 | 0.029 |
| | | 15.73 | 75 | -0.079 | -0.075 | 0.13 | 0.015 |
| | | 16.70 | 40 | -0.10 | -0.12 | 0.12 | 0.020 |
| SDSS MOC (S&C21) | Polana | 12.27 | 6 | -0.027 | -0.022 | 0.038 | 0.015 |
| | | 13.83 | 46 | -0.029 | -0.056 | 0.14 | 0.021 |
| | | 15.12 | 181 | -0.017 | -0.031 | 0.17 | 0.013 |
| | | 16.31 | 122 | -0.007 | -0.034 | 0.19 | 0.017 |
| | Eulalia | 13.11 | 7 | -0.017 | 0.0032 | 0.079 | 0.030 |
| | | 14.65 | 34 | -0.066 | -0.035 | 0.13 | 0.022 |
| | | 15.73 | 110 | -0.051 | -0.059 | 0.15 | 0.014 |
| | | 16.69 | 74 | -0.065 | -0.069 | 0.15 | 0.017 |
| GDR3 | Polana | 11.30 | 4 | -0.046 | -0.022 | 0.040 | 0.020 |
| | | 12.96 | 41 | -0.030 | -0.032 | 0.067 | 0.010 |
| | | 14.35 | 225 | 0.093 | 0.040 | 0.19 | 0.013 |
| | | 15.57 | 91 | 0.17 | 0.17 | 0.18 | 0.019 |
| | Eulalia | 12.99* | 19* | -0.011* | -0.0073* | 0.064* | 0.015* |
| | | 14.30 | 50 | 0.051 | 0.036 | 0.15 | 0.021 |
| | | 15.20 | 86 | 0.10 | 0.076 | 0.17 | 0.018 |
| | | 16.07 | 11 | 0.16 | 0.15 | 0.17 | 0.051 |

**The Eulalia family includes (495) Eulalia in the calculation, but (495) Eulalia was not used as $H_{Min}$, as it would have caused the "largest members" subsample to solely include (495) Eulalia. Instead, the 2nd brightest object (Asteroid 1076) was used as the $H_{Min.}$, and (495) Eulalia was included in the statistical calculations in the largest members' subsample.*

### 3.3 Size-Slope Variability Using a Running Boxcar Method

We employ a running boxcar method to investigate the size-slope variability more finely within the families (Fig. 4). In $H_V$ parameter space (Fig. 4a-c), the Polana family samples are generally characterized by increasing slope with increasing $H_V$ (decreasing size) from the largest sizes until a slope average maximum is reached around $H_V \sim 14.5$ (hereafter, "peak"). For objects with $H_V > 14.5$, the Polana family exhibits a minimum of average slope values (hereafter, "dip") that is most pronounced in the MOC4 and GDR3 samples. The S&C21 Polana family sample shows constant slopes after the slope peak. In (WISE-derived) diameter parameter space (D-space) (Fig. 4d-f), the three Polana family samples show variability in their size-slope behaviors. Increasing slope with decreasing diameter (until the slope peaks are reached) is relatively less steep in D-space for the Polana family's SDSS samples than for the samples in $H_V$-space, though slope dips remain with either size parameter (Fig. 4d-e). The Polana family's MOC4 sample presents a dip with a "flatter" bottom than the "pointed" bottom of its S&C21 sample. The GDR3-derived sample's size-slope behavior in D-space only shows spectral reddening from largest to smallest members (Fig. 4f).

Polana family samples from the SDSS are considered to exhibit relative completeness, but the GDR3 sample is considered incomplete (Table 3). A list of cataloged asteroid family members is considered to exhibit relative completeness down to the size below which object abundance diminishes due to brightness limits. Thus, we determine a sample to have relative completeness when its slope transition size (STS; *e.g.,* peak maximum or dip minimum) occurs at a larger size (smaller $H_V$) than the relative completeness limit (RCL) of the SFD for all cataloged family members (*see:* Table 3 and Thomas et al., 2021).

With the exception of Fig. 4b and Fig. 4f, the Polana family presents a size-slope trend roughly characterized by the “Hygiea-type” from Thomas et al. (2021). In cases where the Polana family shows Hygiea-Type structures, dip minima are between $15 < H_V < 16$ or 4 km < D < 6 km (Fig. 4a-e, Table 3). Although the S&C21 sample does not show the Hygeia-type trend in $H_V$ space, it does in D-space. Similarly, the GDR3 sample shows at least a hint of the Hygeia-type trend in $H_V$ space, even though it does not show it in D-space. Thus, “complete” and “incomplete” Polana family samples show Hygiea-type structure in at least one parameter space.

The Eulalia family samples exhibit more complex relations between spectral slope and size. Both of the Eulalia family's SDSS-derived samples are characterized by fluctuating slope variability. SDSS samples show slopes that increase and decrease through the entire size range of the sample, with both samples showing a slightly more prominent dip around $H_V \sim 15.85$ (Fig. 4a-b, Table 3). The GDR3 sample (in $H_V$ space; Fig. 4c) is characterized by increasing slope with decreasing size until $H_V \approx 15.3$, beyond which decreasing slope with decreasing size is exhibited (Table 3).

When the Eulalia family's SDSS samples are analyzed in D-space, only the MOC4 sample exhibits a flat-bottomed dip from 3 to 5 km (Fig. 4d). The S&C21 sample shows a peak centered around 3.5 km (Fig. 4e), whereas the GDR3 sample in D-space exhibits generally increasing slope with decreasing size, with a potential small peak at ~5 km (Fig. 4f). With the wide variability of size-slope behaviors for the Eulalia family, we are unable to confidently place it within the Hygiea-type or the Themis-type.

There are some qualitative differences between the $H_V$-based and diameter-based analyses of spectral slopes. This variability can be at least partly attributed to variability in albedo among the PEC. Diameter and $H_V$ are related to each other through albedo ($D=1329/pv10-HV/5$). If a constant albedo were assumed, size-slope behaviors would be very similar for either size parameter. However, WISE-derived diameters are computed directly from thermal flux and do not assume a constant albedo. Although our sample lists sorted by $H_V$ and diameter have considerable overlap, they are not ordered exactly the same, leading to slight differences between the respective plots. Notably, the transition from increasing to decreasing spectrophotometric slope (or *vice versa*) mostly remains in both $H_V$- and D-spaces for the SDSS samples, suggesting that $H_V$ is a reliable proxy for size in size-slope analyses.

There are also quantitative differences between the SDSS and GDR3 samples. Although the slope values for both families in each of the three catalogs are on the same order of magnitude at all sizes, the families show less overlap in slope-space for the SDSS samples (Fig. 4a-b,d-e) than in the GDR3 samples (Fig. 4c,f), especially at smaller (< 9 km) sizes. In addition, the GDR3 samples show much redder slopes than the SDSS samples. Galinier et al. (2023) note that Gaia spectra tend to have redder slopes than spectrophotometry of the same objects in SDSS. Although no cause is specified, they suggest differences in the solar standards used in the datasets could be one possibility.

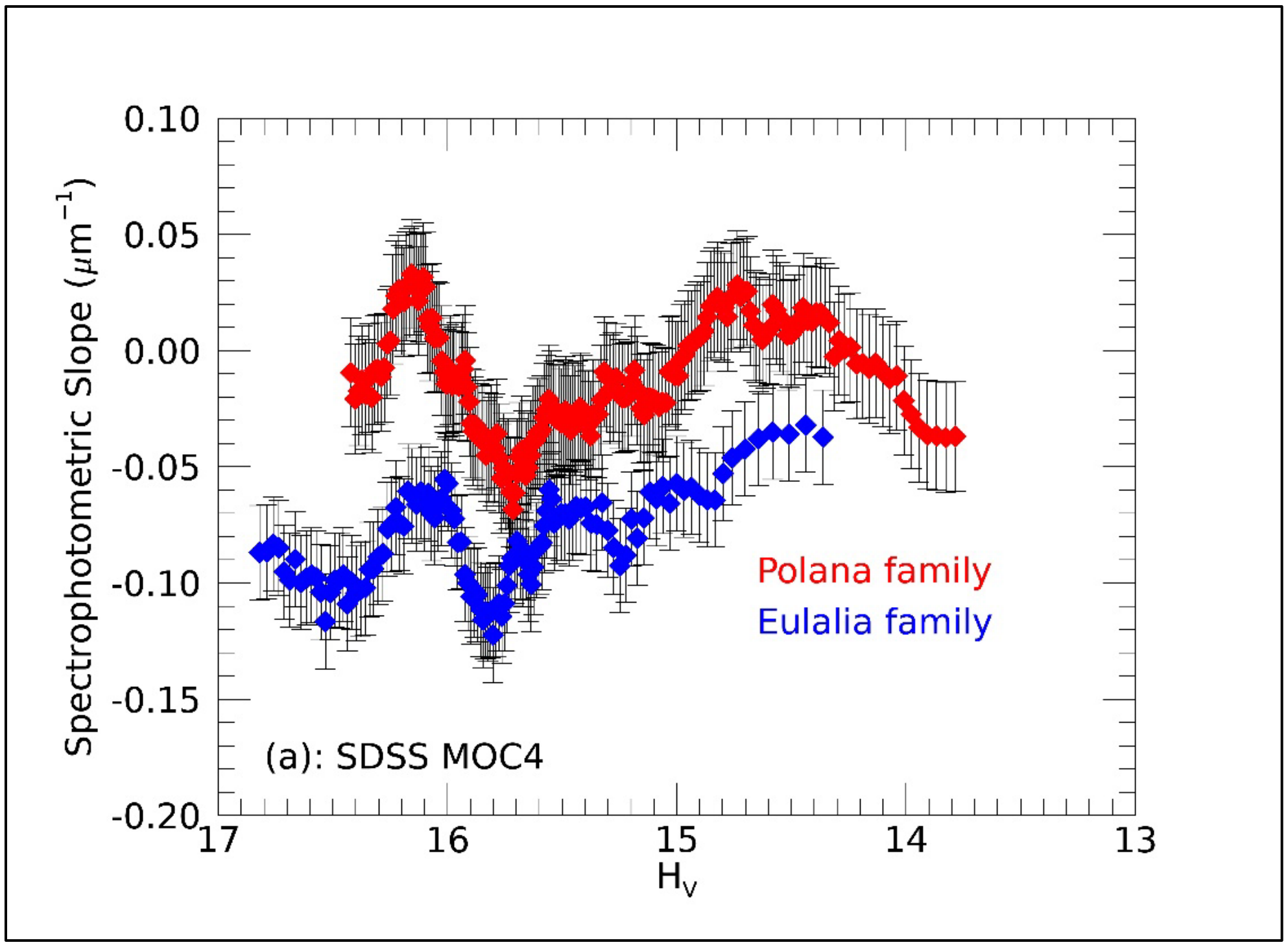

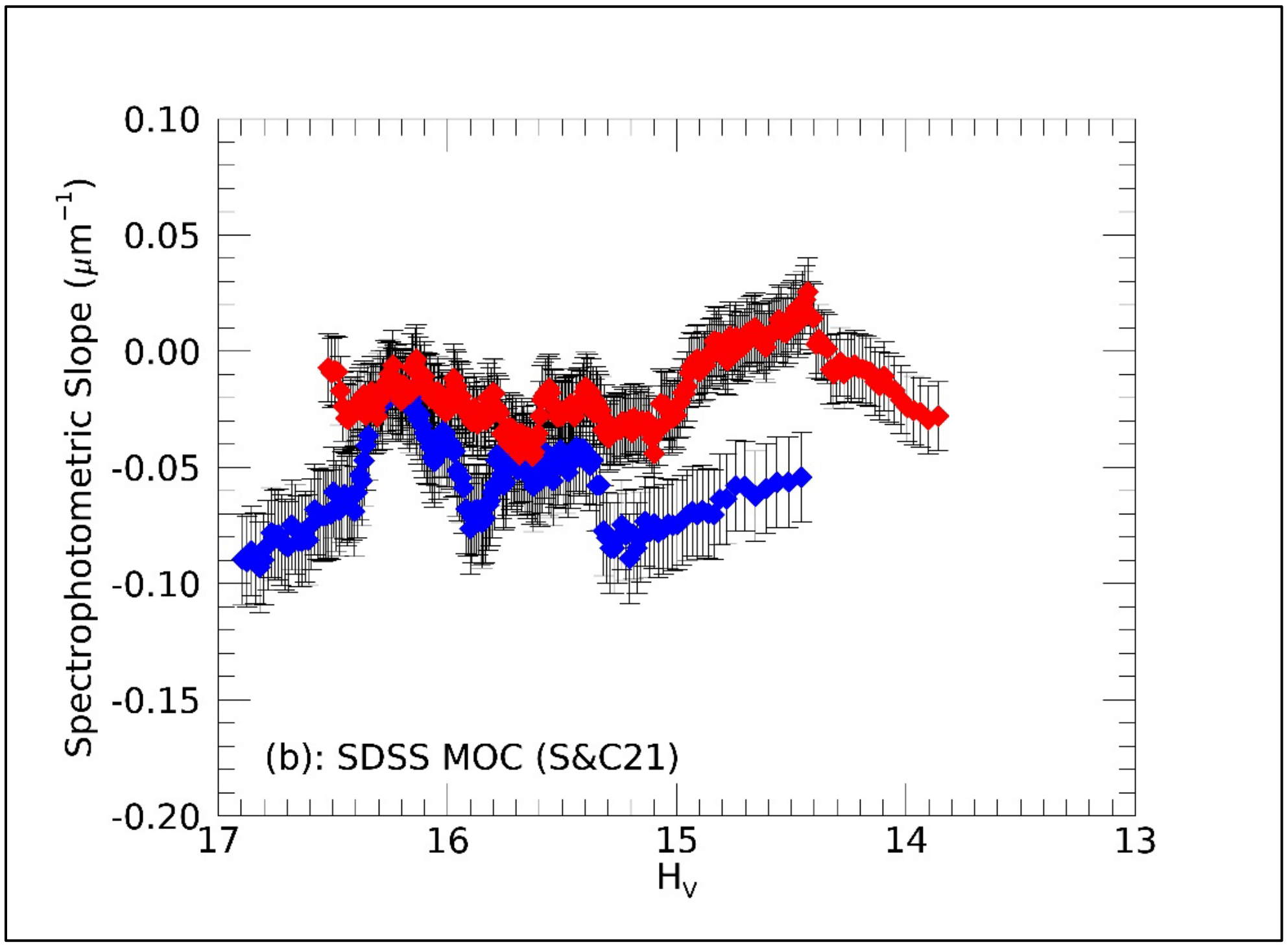

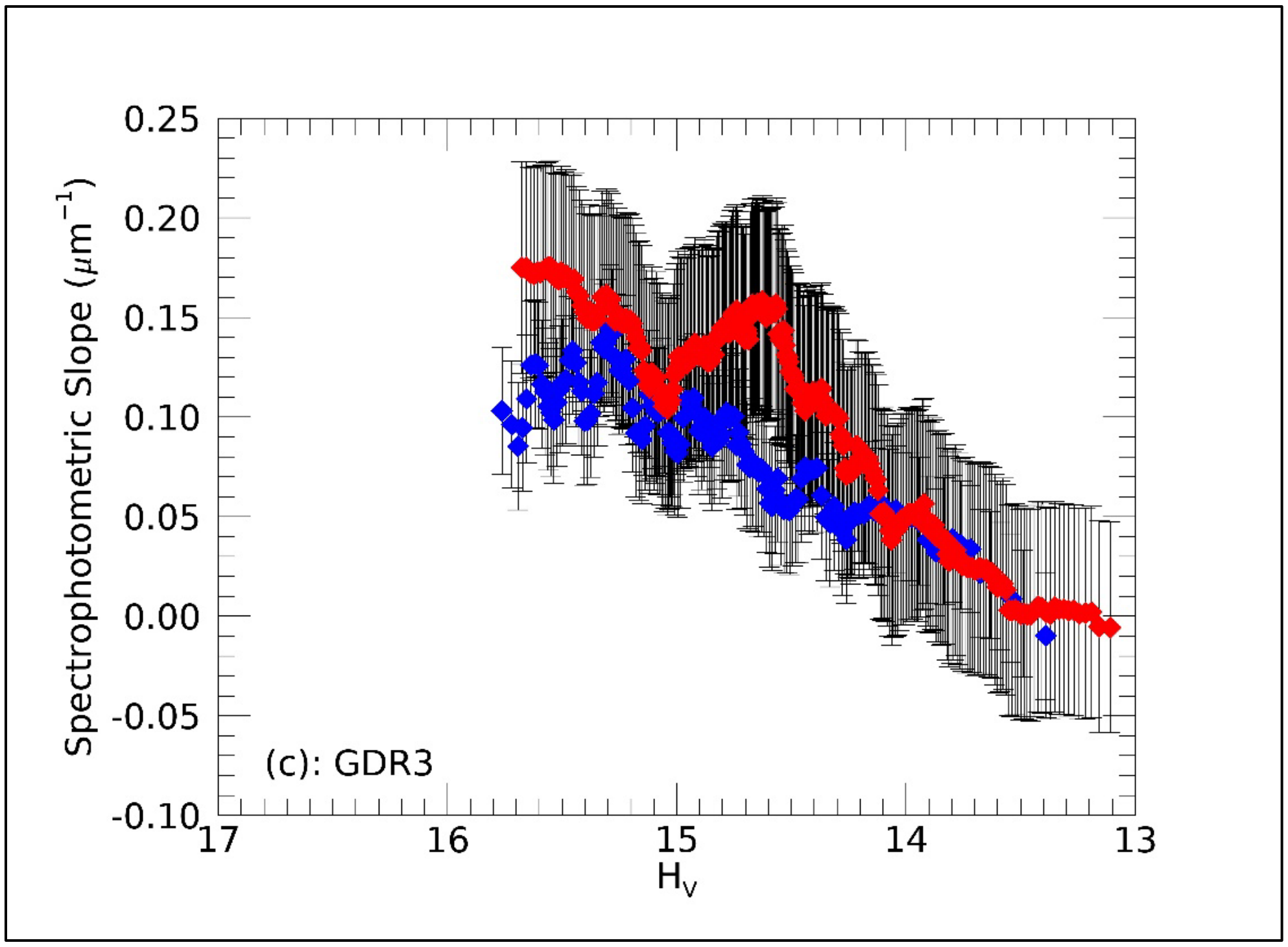

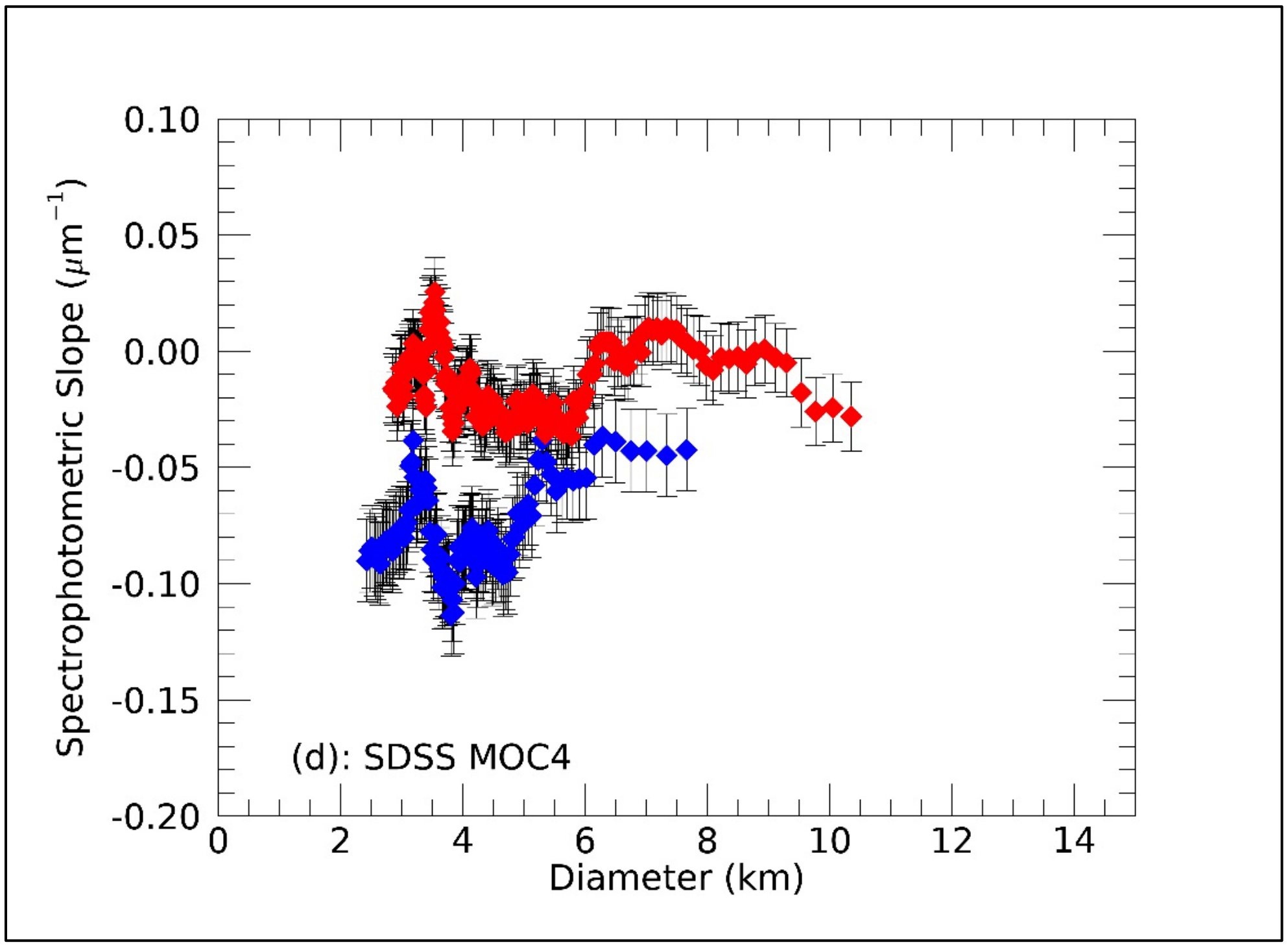

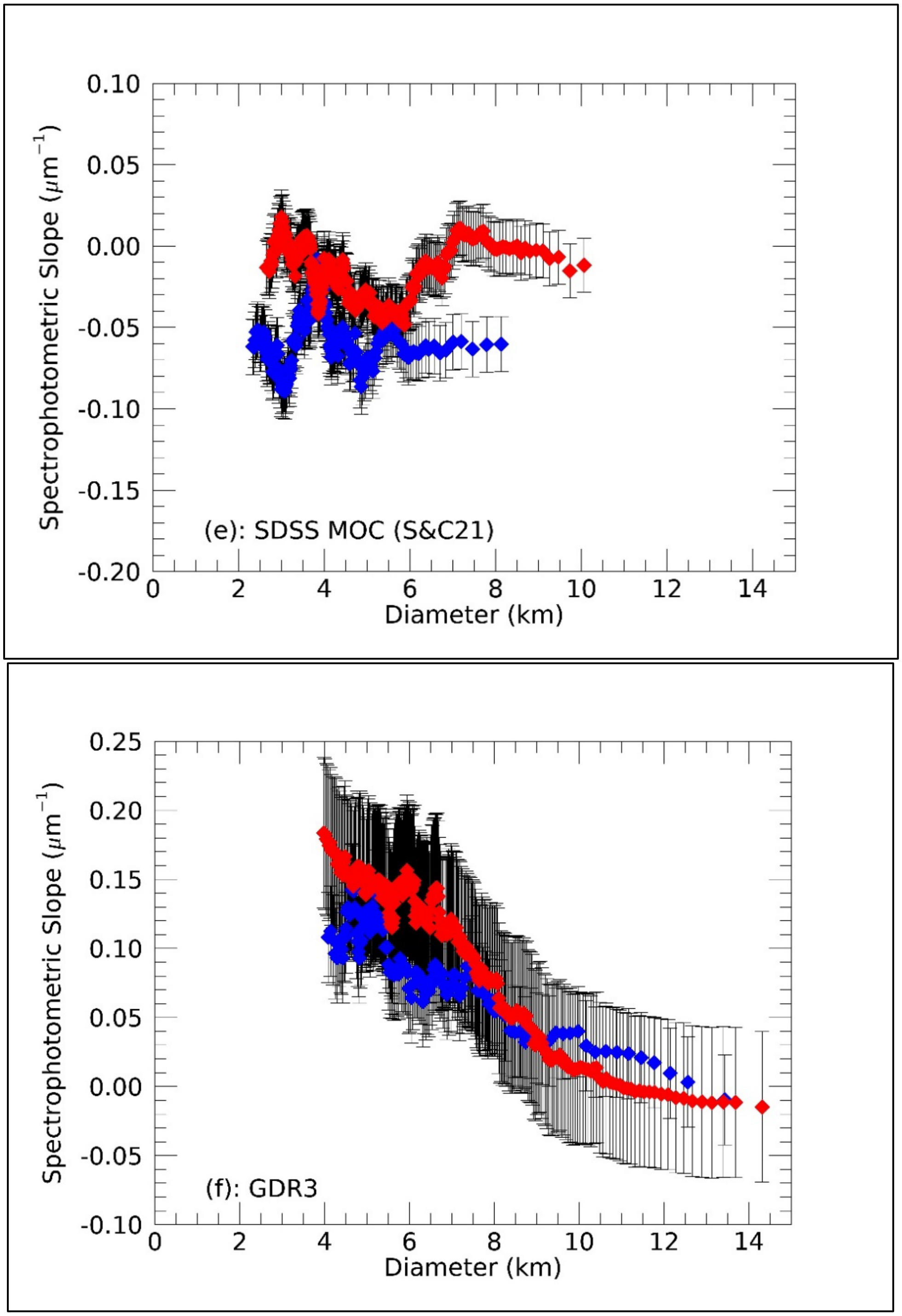

**Figure 4**: *Spectrophotometric slope (μm$^{-1}$) as a function of size for both PEC families, as observed through all the samples extracted from the SDSS MOC4 (a & d), the SDSS MOC release form Sergeyev & Carry (2021) (b & e), and the GDR3 (c & f).*

| Family | Catalog | Relative Completeness | Slope Transition | Relative Completeness |
|---|---|---|---|---|

| | | | Size ($H_V$) | Limit ($H_V$) |
|---|---|---|---|---|
| Polana family | SDSS MOC4 | Marginally Complete [a] | 15.71 | 15.80 |
| | SDSS MOC (S&C21) | Complete | 15.10 | 16.20 |
| | GDR3 | Incomplete | 15.04 | 14.80 |
| Eulalia family | SDSS MOC4 | Complete | 15.80 | 16.00 |
| | SDSS MOC (S&C21) | Marginally Complete | 15.90 | 16.00 |
| | GDR3 | Complete | 15.29 | 15.60 |

**Table 3:** The completeness of each family for each dataset.
*Completeness is defined by the slope transition point and the SFD of the family. If a family's slope transition size (STS) is greater than the relative completeness limit (RCL), then the family is considered complete (i.e., STS > RCL). [a]Marginal completeness refers to instances when completeness is when |STS – RCL| ≤ 0.1-mag from the completeness limit.*

# 4 Discussion

## 4.1 Visible Spectral Differences between the Polana and Eulalia families

Our work finds that the Polana family is marginally redder than the Eulalia family in VIS wavelengths, which agrees with prior work. Tatsumi et al. (2022) showed slightly redder NUV-VIS slopes for the Polana family than for the Eulalia family. Delbo et al. (2023) showed statistically significant redder NUV slopes for the Polana family than for the Eulalia family. Ground-based spectra of PEC members found the average slope for the Eulalia family to be slightly redder than for the Polana family (de León et al., 2016; Pinilla-Alonso et al., 2016). When we compute VIS spectral slope from each ground-based spectrum to match the wavelength range investigated in this work (and normalized to the reflectance at λ = 0.62 μm), we find consistency with our work, with the Polana family ($S' \approx 0.047 +/- 0.03\ \mu m^{-1}$) being marginally – though not significantly – redder than the Eulalia family ($S' \approx 0.038 +/- 0.02\ \mu m^{-1}$). We note that the slopes in our work are calculated from ~0.47–0.89 μm, whereas slopes were calculated from 0.55 to 0.9 μm in de León et al. (2016), reinforcing the findings of Delbo et al. (2023) that slope difference between the two PEC families manifests at the shorter wavelengths.

Our results show that the significance of the VIS slope distinguishability for the families depends on the dataset. The SDSS-reported slopes of the Polana and Eulalia families show a low probability of having originated from the same population. The GDR3 samples do not support significant distinguishability in VIS slope space (Table 1), though we note that the Polana family's GDR3 sample is considered incomplete (Table 3). Insignificant distinguishability between the PEC families in the GDR3 samples actually agrees with Delbo et al. (2023), in that they found that the PEC families show spectral distinction only in the NUV wavelengths – not VIS wavelengths. We note that VIS indistinguishability for the GDR3 samples may be related to the currently-elusive cause behind the redder slopes of *Gaia* spectra (*see:* Galinier et al., 2023).

The size range for the SDSS samples and the GDR3 samples skew toward smaller and larger objects, respectively. Thus, the significance discrepancy between the SDSS and GDR3 samples (Table 1) potentially suggests a size-dependence to the distinguishability of the PEC families, with the smaller members being more distinguishable than their larger counterparts. More data (such as from the Vera C. Rubin Observatory, VCRO) are necessary to fully establish whether there is a size-dependence to the distinguishability between the PEC families.

**4.2 Insights from Small PEC Members**

One major theme of our work is that VIS spectral slope varies with size for PEC members, which provides more support to size-slope behaviors within C-complex populations generally. As shown in Fig. 3, smaller PEC members often show greater VIS slope variability than the larger members. The running box averages (Fig. 4) also show variability, particularly around the smaller-size end of each of the samples ($\lesssim$ 6-km). Our results are therefore consistent with previous studies of other asteroid groups that show size-slope behaviors bifurcate large and small asteroids by VIS spectral slope (*e.g.,* Binzel et al., 2004; Thomas et al., 2012; Beck and Poch, 2021; Thomas et al., 2021).

Our work has potential connections to Bennu and Ryugu, particularly because of their sub-km sizes. VIS spectral slopes of Bennu (Clark et al., 2011; Hergenrother et al., 2013; Hamilton et al., 2019; Lauretta et al., 2024) and Ryugu (Vilas, 2008; Kitazato et al., 2019) are within the range of slopes reported in this work for each of the three datasets, though we note the sub-km sizes of Bennu and Ryugu are outside the size range of our catalog-derived PEC samples. In this sense, our work agrees with de León et al. (2016), which also found that the PEC's VIS slopes span a range that includes those of Bennu's and Ryugu's. Potentially consistent with dynamical models of Bennu and Ryugu (Campins et al., 2010, Campins et al., 2013; Bottke et al., 2015), the wider slope distribution(s) of the Polana family may correspond to the higher dynamical probability that Bennu and Ryugu originate from the Polana family (~70 %) than from the Eulalia Family (~30 %) (*e.g.*, Bottke et al., 2015). Alternatively, the PEC's relative average slopes may be indicating more likely origins for Bennu (blue-sloped upper surface) and Ryugu (red-sloped) to be the (bluer) Eulalia and (redder) Polana families, respectively, consistent with NUV-based arguments by Delbo et al. (2023).

**4.3 Potential Compositional Differences Between the Polana & Eulalia Families**

Bulk, VIS slope differences between the PEC families may be suggestive of bulk, physical differences between the families. Additionally, comparative analysis of two similar families, like those of the PEC, may be a way to determine (or at least constrain) cause(s) of size-slope behaviors in C-complex populations generally. While the leading hypothesis to explain the analogous size-slope trends for S-complex populations invokes space weathering (SW) and re-surfacing mechanisms (*e.g.*, Binzel et al., 2004; Thomas et al., 2012; Graves et al., 2018; Vokrouhlický et al., 2015), Thomas et al. (2021) hypothesized these mechanisms may act alongside other factors for size-slope trends of C-complex populations. We briefly explore one hypothesis for the bulk observed differences between PEC families: hydration state.

The bulk slope difference may be indicating that the Eulalia family is more hydrous than the Polana family. In laboratory-based experiments that simulate SW environments, carbonaceous chondrites (CCs) that are hydrous (CI and CM), generally respond to irradiation and micrometeoroid bombardment with spectral bluing, whereas anhydrous CCs (CV/CO/CK) respond to SW environments with spectral reddening (Hiroi et al., 2013; Keller et al., 2015; Lantz et al., 2015; Lantz et al., 2017; see Trang et al., 2021 for a review). Therefore, bluer slopes of the Eulalia family may be implying a higher abundance of hydrous material compared to the Polana family. Characterization at longer wavelengths is necessary to determine the PEC's hydration status.

The hydration state of the PEC remains unknown. No observed PEC member exhibits a 0.7-μm feature associated with Fe-bearing phyllosilicates (de León et al., 2016), so observations of the 3-μm region, which covers the fundamental OH vibrational mode, are necessary to fully determine the hydration status of the PEC (Rivkin et al., 2002). Such IR (2.2–5 μm) observations are sparse for the PEC. The only ground-based observation of the 3-μm region for a PEC member was conducted for (142) Polana by Takir et al. (2024), which did not find the presence of a strong 3-μm feature. The JWST Cycle 2 Program #3760 has observed (142) Polana with NIRSpec, so the community will be further informed of its hydration state upon the publication of its JWST spectrum that is un-obstructed by telluric contamination.

Both Bennu and Ryugu exhibit absorptions in the 3-μm region that are suggestive of Mg-rich phyllosilicates (Kitazato et al., 2019; Hamilton et al., 2019; Lauretta et al., 2024). However, Bennu (blue) and Ryugu (red) also show different VIS slopes, despite both being hydrated. Differences in their band depths and centers may indicate different degrees of aqueous alteration (*see:* Takir et al., 2013). Band depths from both remote sensing and returned samples from Bennu and Ryugu are ~ 20 % (Hamilton et al., 2019; Lauretta et al., 2024) and 12–18 % (Kitazato et al., 2019*;* Pilorget et al., *2022*), respectively. Remote sensing observations showed Bennu and Ryugu exhibit band centers around 2.74 μm (Hamilton et al., 2019) and 2.72 μm (Kitazato et al., *2019*), respectively, whereas returned samples showed centers at ~2.7 μm. Thus, spectral comparisons between CCs and remote observations point to CM-like and CI-like compositions for Bennu and Ryugu, respectively (Hamilton et al., 2019; Kitazato et al., *2019*). However, returned samples from both Ryugu and Bennu's Hokioi Crater are suggestive of CI-like composition for both NEAs (Pilorget et al., 2022; Yokoyama et al., 2023; Lauretta et al., 2024). Additionally, Yumoto et al. *(2024)* found that VIS slopes of fresh craters across Bennu's and Ryugu's surfaces are similar, but space weathering trends appear to bluen Bennu's bulk surface and redden Ryugu's, which is potentially attributable to grain size differences between the two NEAs. The slope variations induced by space weathering on Bennu and Ryugu are comparable to our work's slope distributions for the Eulalia and Polana family samples, respectively. With the IR context of Bennu and Ryugu, we maintain that the PEC's bulk slope distributions may be indicative of the relative hydration states of the families, but the VIS slope of an individual member, like Bennu or Ryugu, is not a reliable proxy for its specific hydration status.

More observations of PEC members – large and small – will be necessary to fully characterize any size-dependent nature to the PEC's hydration state. Fortunately,

JWST's Cycle 3 GO schedule includes NIRSpec observations to investigate the 3-μm region for (495) Eulalia and three, ~10-km-sized Polana family members (Program #6384). Three scenarios could result from both aforementioned JWST observations. (I) Neither PEC family presents a 3-μm absorption, and our VIS-based results may not indicate anything about hydration state for the PEC. (II) None of the Polana family asteroids exhibits a 3-μm absorption, but (495) Eulalia does, perhaps implying that our bulk averages of VIS slope act as proxy for hydration status of the families. (III) Neither of the largest members of the families exhibit 3-μm features, but the smaller Polana family members do, suggesting a size-dependence to hydration state. A size-dependent compositional trend for the PEC is an exciting possibility that would provide insight and more questions about the parent bodies of these families. We note additional JWST observations of small Eulalia family members would be a vital complement to understanding the hydration state of the family.

We note that bulk slope differences between the PEC families may also be partly related to other compounding factors. Polana family asteroids are redder, perhaps, because they exhibit smaller average grain sizes than Eulalia family asteroids (*see:* Vernazza et al., 2016; Cantillo et al., 2023). Additionally, the narrower slope distribution of the Eulalia family could be related to its younger age relative to the Polana family (see Fig. 1; **Section 1.2**; Walsh et al., 2013). Bluer Eulalia family slopes could potentially reflect “fresher” regolith than compared to weathered, older Polana family slopes. However, SW timescales likely operate on relatively short timescales, much shorter than either family's age (Walsh et al., 2013; Graves et al., 2018; Thomas et al., 2021).

### 4.4 The PEC & Morate Cone Structure

Morate et al. (2019) found that the apparent spread of spectral slopes among the primitive IMB population increased with decreasing size. They fit lines to average $H_V$ values in 4 equal-sized bins and ± 1σ (and ± 2σ) variations in each bin, which we again refer to here as Morate Cones. We fit lines (equations provided in **Appendix's** Table A) between the average $H_V$ and ± 1σ/±2σ for size bins data in Table 2 for both families (Fig. 5a-c). We exclude (495) Eulalia from the calculations that derive its Morate Cone's lines in the GDR3 sample (Fig. 5c). (495) Eulalia is considerably larger than the rest of its GDR3 sample, effectively making it act as an outlier that affects the calculation of its ±1σ and ± 2σ lines if it is included.

Table 2 demonstrates that the distributions often overlap and are unequally distributed in sample size. In some instances, we do not find increasing standard deviation with decreasing size, perhaps a byproduct of small sample sizes within some of the bins that in turn make conducting our Morate Cone analyses unreliable currently. Larger survey catalogs (*e.g.,* from the VCRO) might allow for greater sample sizes for each of these bins. Until then, Morate Cones for the PEC remain preliminary.

Our results, combined with the previous literature (Morate et al., 2019; Tinaut-Ruano et al., 2024) may suggest that the Morate Cone is a property of specific populations within the MB and not a bulk property of the region. That is, the original Morate Cone was formed from a suite of IMB, C-complex populations, notably not just one family (Morate et al., 2019). When Tinaut-Ruano et al. (2024) used GDR3 data to expand the investigation of size-slope spread to the entire MB's primitive asteroid population, they found no substantial difference of slope ranges among various sizes.

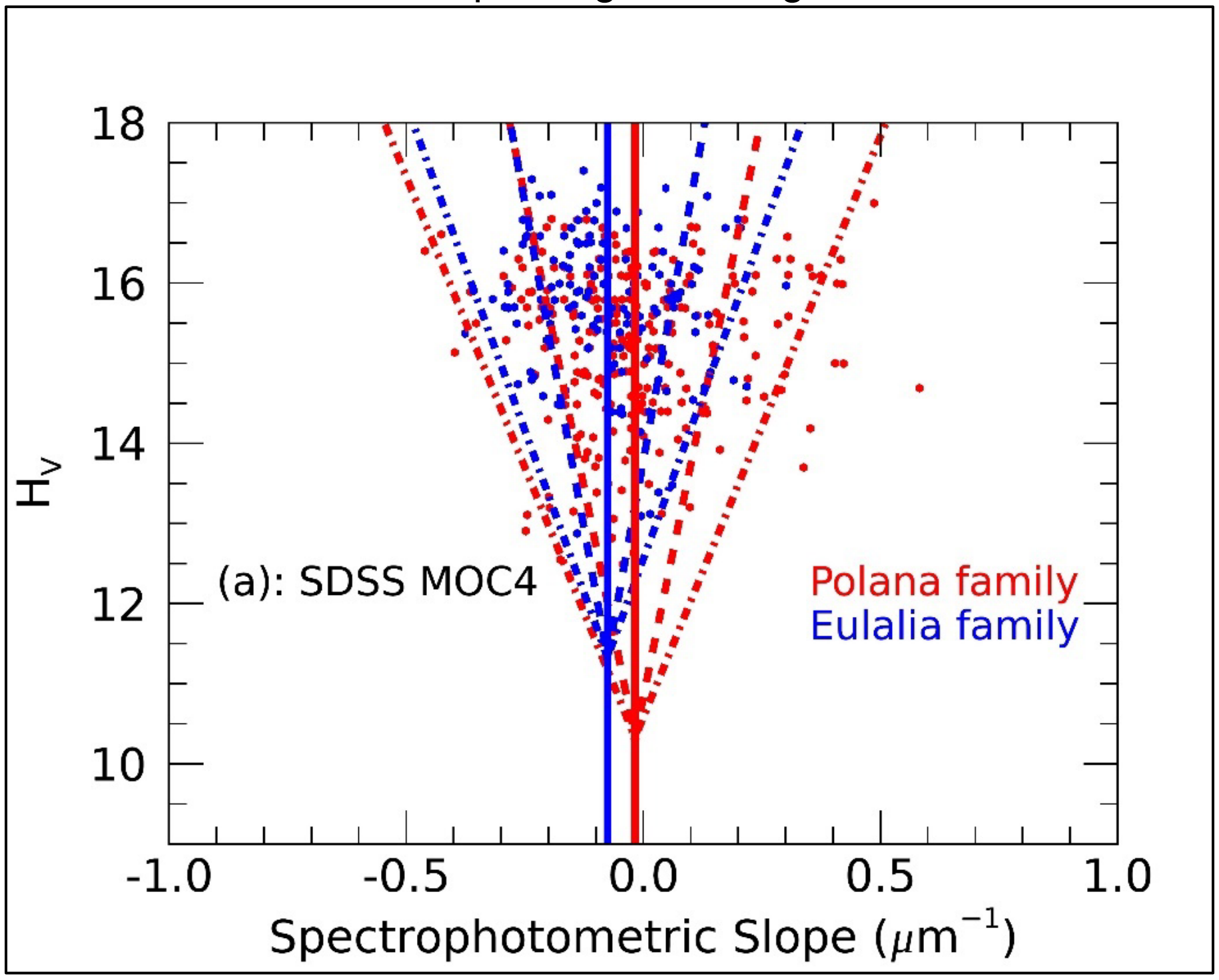

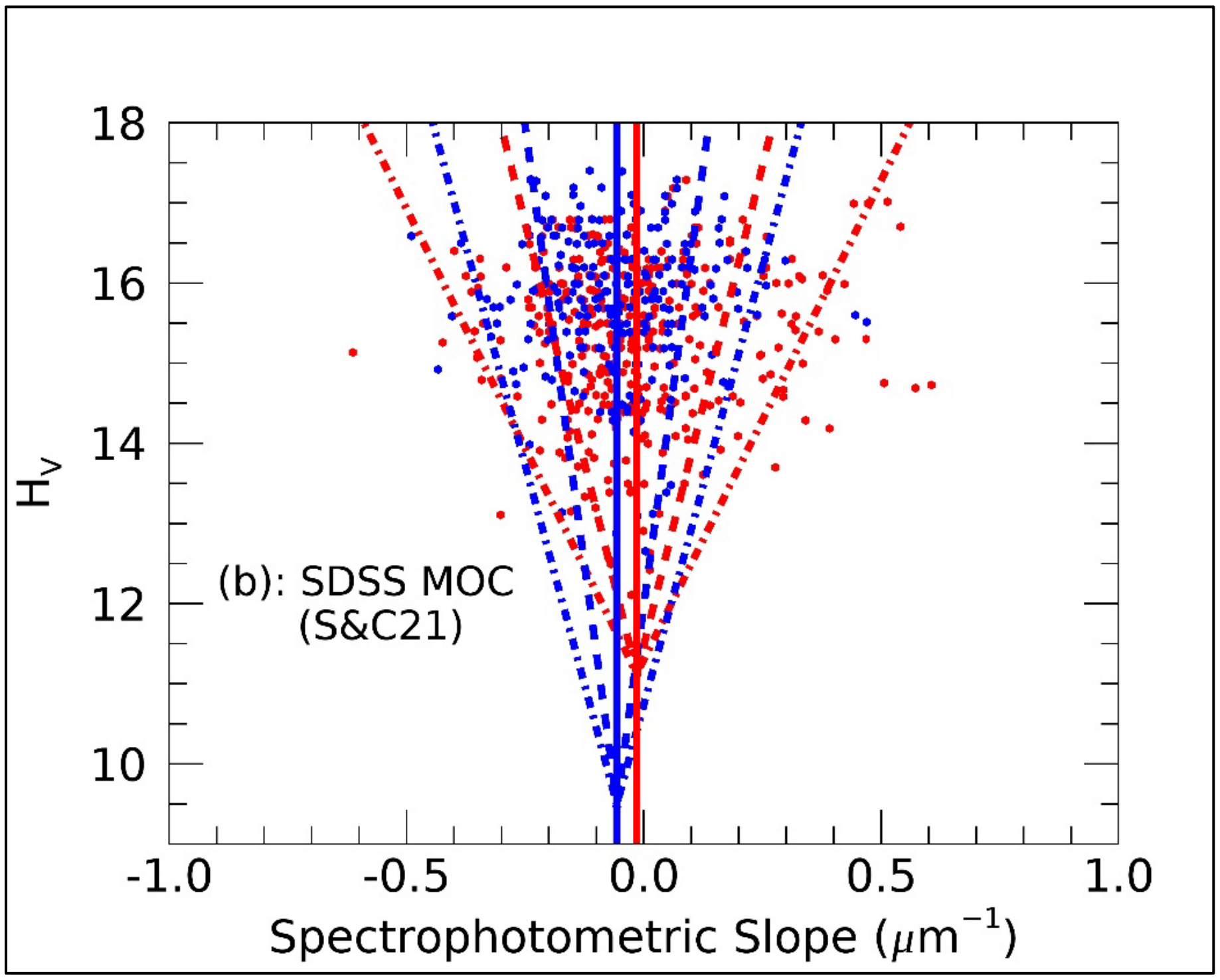

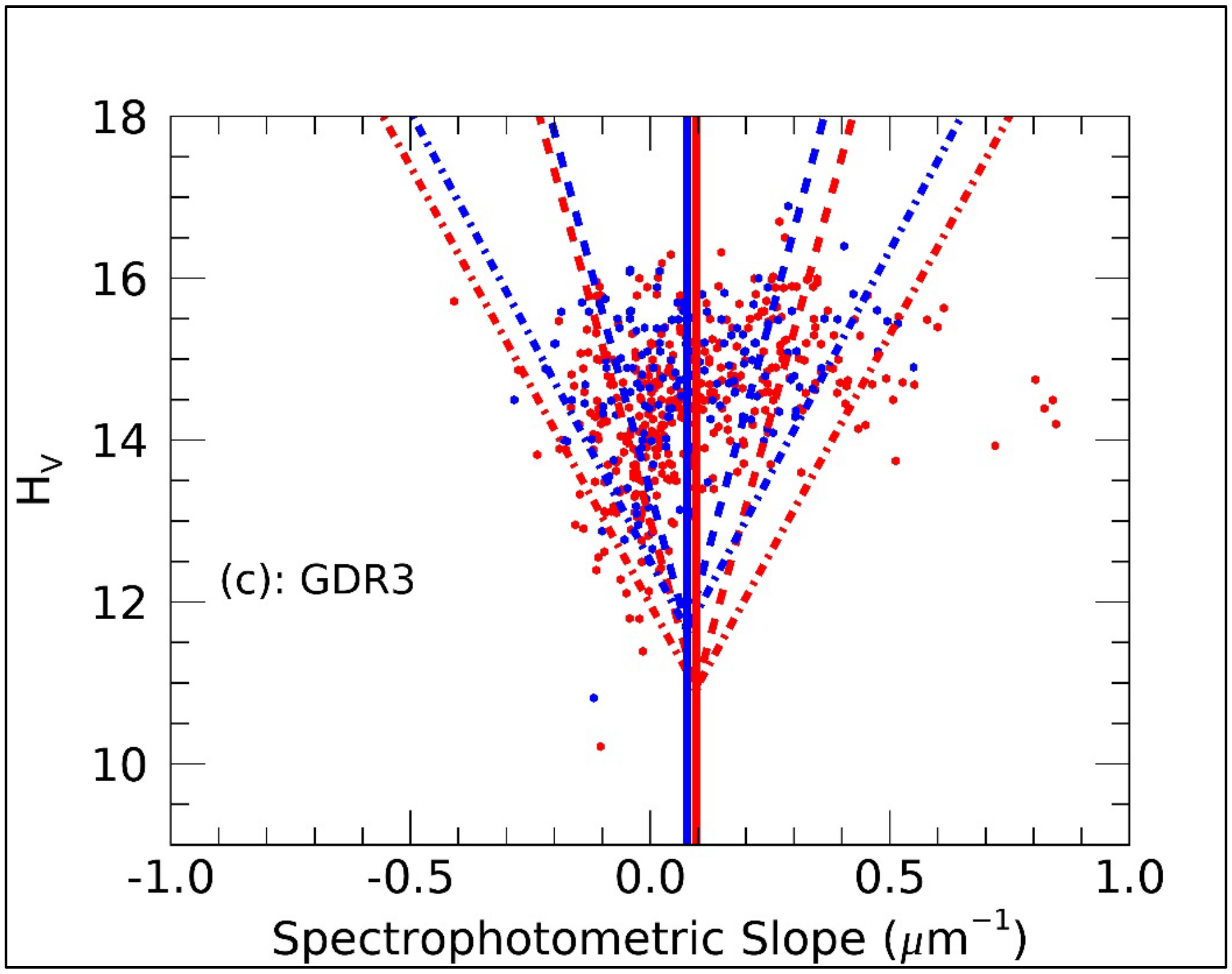

**Figure 5**: *Spectral slope vs absolute magnitude for all members of the Polana (red) and Eulalia (blue) families in all three catalogs. Though preliminary, see a similar structure as reported in Morate et al. (2019) for other inner Main Belt asteroids (i.e., Morate Cones). The solid vertical lines represent the average slope of the family, while the dashed, diagonal dashed and dash-dotted lines represent the 1σ and 2σ spreads, respectively, from the average for the 4 size bins (**Table 3**). The equations for the dashed lines are provided in the **Appendix's Table A**.*

**4.5 Size-Slope Trend Reliability**

The GDR3 data of the Polana and Eulalia families appear to show different size-slope behaviors from the SDSS data. To test whether this difference is common, we used the GDR3 catalog to characterize size-slope behaviors for the Hygiea and Themis families (the large MB families of the prototypical "types" in Thomas et al., 2021). We chose to compare the GDR3 with just the MOC4 because we observed closer agreement of slopes/size-slope behaviors between the two SDSS releases used in this work and because Thomas et al. (2021) used MOC4. We used the Hygiea and Themis family lists from Nesvorný, 2015 to extract members with low errors (< 8 %) in all filters, no flag in any filter (from 0.462 to 0.902 μm), |C| < 1, and − 0.2 < $R_z$ - $R_i$ < 0.185, as we had also done for the PEC families. The Hygiea and Themis family lists do not have known, S-type interlopers from an overlapping family, as is the case for the Nysa family's contamination in the PEC family lists (Walsh et al., 2013). Also, unlike the PEC families, albedo measurements do not exist for the entire extracted Hygiea and Themis family samples from GDR3 (Mainzer et al., 2016). For these reasons, we do not invoke the additional selection criterion of albedo for these families.

The Hygiea (*N* = 278; Fig. 6a) and Themis (*N* = 878; Fig. 6b) families show differences between their SDSS and GDR3 samples. For the Hygiea family, the GDR3 sample shows a dip, but the center and depth of the dip is different between the two samples. For the Themis family, the GDR3 sample shows a monotonic increase in spectral slope with decreasing size (increasing $H_V$). The size range covered by the Themis family's GDR3 sample roughly corresponds to the larger-size half of the size regime covered by the SDSS sample. Thus, the GDR3 Themis family sample may be showing a portion of the full size-slope trend due to a lack of small members. Additionally, slopes from both GDR3 samples are redder than from their respective SDSS samples, due likely to the slope differences previously observed between the two datasets (*e.g.,* Galinier et al., 2023).

The dataset-dependent size-slope behaviors for the Hygiea and Themis families is particularly relevant to observed differences between the PEC families' behaviors in Fig. 4. The relative redness of the GDR3 samples is consistent with what is shown for the PEC families (Table 1; Fig. 4). The potential “partial view” of the size-slope trend for the Themis family may suggest an inadequate number of small PEC members that affects the size-slope behavior exhibited by the PEC's GDR3 samples. Altogether, dataset-dependent differences underscore the necessity of small member inclusion and of refinement to the definition of “relative completeness.”

We caution against relying too heavily on size-slope trend types, despite the prevalence of size-slope trends in Thomas et al. (2021) and this work. The Polana family shows decent consistency with the Hygiea-Type even with our stricter data collection limits. However, the Eulalia family, whose data selection was mostly affected by the de-selection of potential dynamical interlopers (a factor also included in Thomas et al., 2021), does not resemble either the Hygeia- or Themis-type trends. We reiterate that the eventual creation of a moving object catalog from the VCRO will be vital to a better understanding of size-slope trends in these families.

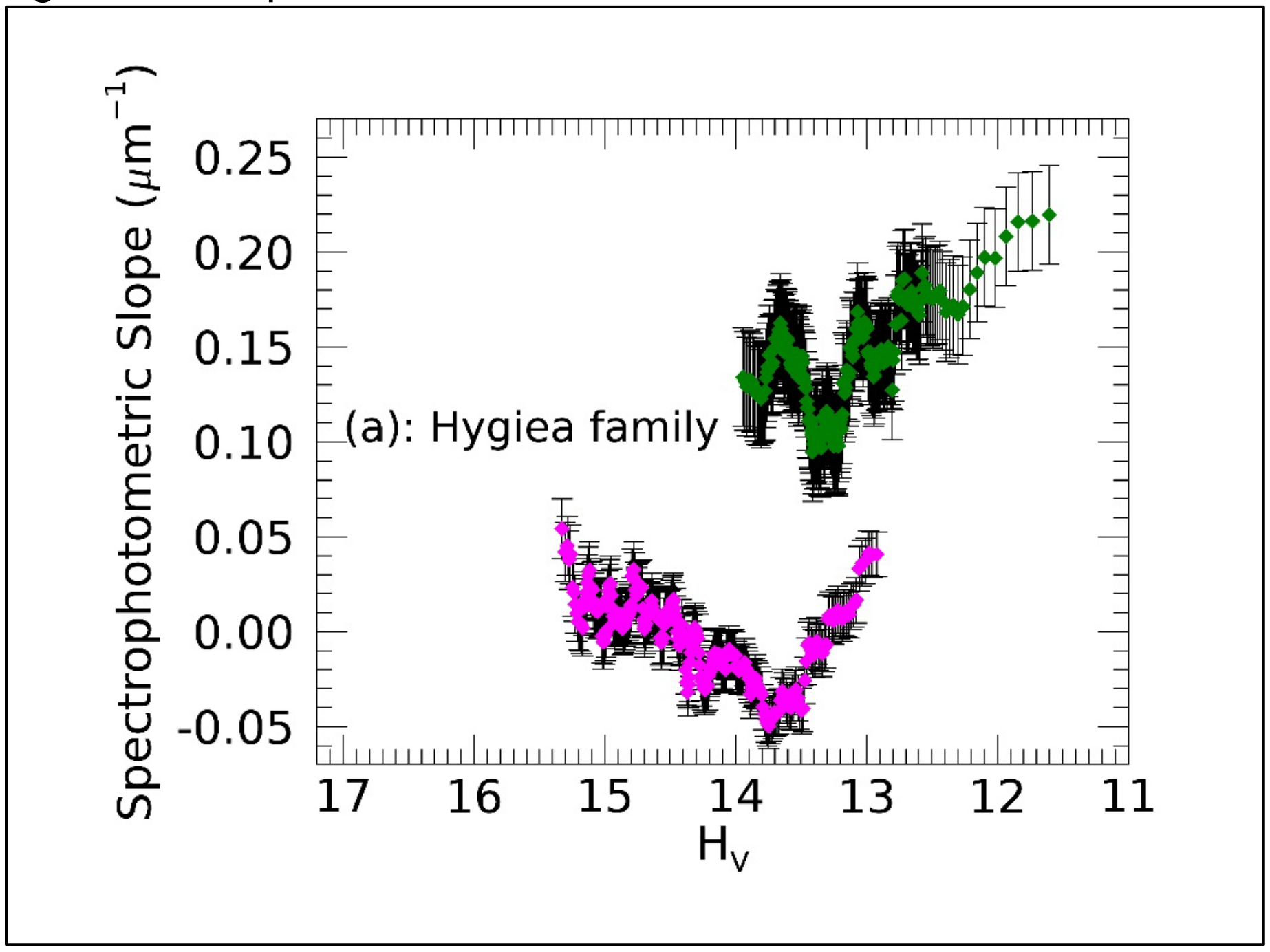

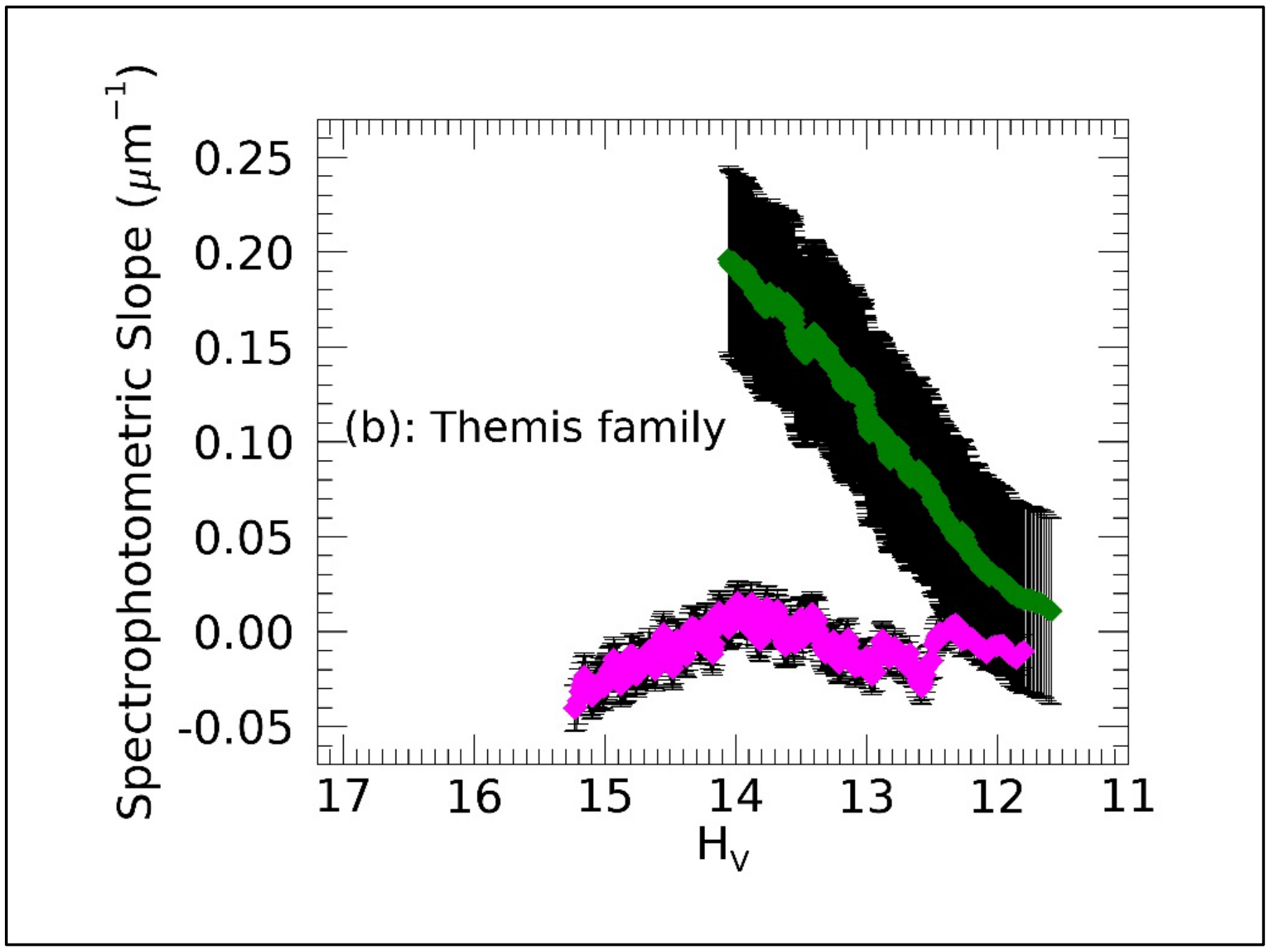

**Figure 6**: *Size-slope trends for the (a) Hygiea and (b) Themis families. The GDR3-derived samples are green, while the SDSS MOC4 samples from Thomas et al. (2021) are in magenta, with both size-slope trend errors shown in black.*

## 5 Conclusions

Our work shows the Polana family is generally more positively-sloped (redder) than the Eulalia family in VIS wavelengths. The relative redness of the Polana family to the Eulalia family extends support to the findings from the NUV-based work from Delbo et al. (2023) (and their re-analysis of Tatsumi et al. 2022) that the two families are spectrally distinguishable. Although ground-based, VIS spectra (de León et al., 2016) showed slightly bluer slopes for the Polana family, re-calculations of spectral slopes to the wavelength range (and normalization wavelength) used in our work show further consistency with our main result. GDR3 bulk comparisons suggest, however, that the statistical significance of VIS distinguishability may depend on object size, with smaller objects being more distinguishable than larger ones. More data are needed to investigate that claim more fully.

Our results provide potential insights about Bennu and Ryugu and their connection to the PEC. Both NEAs' VIS slopes fit within either family's slope distribution. Consistent with Delbo et al. (2023), the Eulalia family's average spectral slope agrees more with the VIS slopes measured for Bennu's upper surface layer (though maybe not the subsurface, *see:* Lauretta et al., 2024), whereas the Polana family's average slope is more similar to the VIS slopes measured for Ryugu.

Slope differences in our work may be suggestive of bulk differences between the PEC families, like hydration state. In particular, the bluer slopes of the Eulalia family may be implicating a higher occurrence of the 3-μm feature for the family. Observations from two JWST observing campaigns will be able to fully demonstrate if our suggestions are correct or if other factors (like grain size and dynamics) are more responsible for the observed distributions.

Size-slope trends from Morate et al. (2019) (*i.e.*, “Morate Cones”) are observed for both PEC families. Our work's implementation of Morate Cones underscores that it is potentially a valuable method for distinguishing slopes between extremely similar asteroid families, but we note limitations to this method, specifically when small subsample sizes are present (see Table 2). Our work, combined with Morate et al. (2019) and Tinaut-Ruano et al. (2024), may be suggesting that populations within the Main Belt – though not the Main Belt itself – express this kind of size-slope variability.

The size-slope trend types from Thomas et al. (2021) are exhibited by a portion of our samples, but there is a dependence on the catalog and size parameter employed. We provide evidence that the Hygiea and Themis families themselves show discrepancy between the MOC4 and GDR3 datasets, similarly to the differences observed for the PEC families. We attribute these differences partly to each type's sensitivity to the analyses performed to generate the size-slope trend. Nevertheless, slope variation with size that is observed by both families suggests some size-dependent control on physical properties, perhaps analogous to those previously discussed among the S-complex.

## Acknowledgements

We would like to thank Noemí Pinilla-Alonso and Mário De Prá for helpful insights at the beginning of this work. We would also like to thank Max Mahlke for the recommendation to implement the later release of the SDSS MOC from Sergeyev & Carry (2021), which made this work more comprehensive. We also thank Marco Delbo, Chrysa Avdellidou, and Marjorie Galinier for their helpful aid in Gaia-related portions of this work. This work was funded through a Future Investigators in NASA Earth and Space Science and Technology (FINESST) research grant (solicitation: ROSES-2023 NNH22ZDA001N-FINESST).

## Appendix:

**Table A**: The +/-1σ and +/-2σ lines for the Morate Cones in **Fig. 4a-c.**

| The Polana family | | The Eulalia family | |
|---|---|---|---|
| *SDSS MOC 4* | | | |
| +1σ | H = 29.02(S') + 10.84 | +1σ | H = 32.50(S') + 13.78 |
| -1σ | H = -29.02(S') + 9.821 | -1σ | H = -32.50(S') + 8.873 |
| +2σ | H = 14.51(S') + 10.58 | +2σ | H = 16.25(S') + 12.56 |
| -2σ | H = -14.51(S') + 10.07 | -2σ | H = -16.25(S') + 10.10 |
| *SDSS MOC (Sergeyev & Carry 2021)* | | | |
| +1σ | H = 23.93(S') + 11.48 | +1σ | H = 43.82(S') + 11.99 |
| -1σ | H = -23.93(S') + 10.77 | -1σ | H = -43.82(S') + 7.007 |
| +2σ | H = 11.96(S') + 11.30 | +2σ | H = 21.91(S') + 10.74 |
| -2σ | H = -11.96(S') + 10.95 | -2σ | H = -21.91(S') + 8.252 |
| *GDR3* | | | |
| +1σ | H = 21.63(S') + 8.859 | +1σ | *H = 22.20(S') + 9.939 |
| -1σ | H = -21.63(S') + 13.006 | -1σ | *H = -22.20(S') + 13.36 |
| +2σ | H = 10.81(S') + 9.895 | +2σ | *H = 11.10(S') + 10.79 |
| -2σ | H = -10.81(S') + 11.97 | -2σ | *H = -11.10(S') + 12.51 |

**Instances where the largest object (i.e., 495 Eulalia) are excluded from calculation of the $H_V$ bins due to a dramatic difference in size compared to the rest of the sample.*